\documentclass[aps,11pt,preprint,superscriptaddress]{revtex4}
\usepackage[usenames]{color}
\usepackage[dvips]{graphicx}
\usepackage{amsmath}
\usepackage{amsfonts}
\usepackage{amssymb}
\usepackage{mathrsfs}
\usepackage{bm}
\usepackage{verbatim}
\newcommand{\beq}{\begin{equation}}
\newcommand{\eeq}{\end{equation}}
\newcommand{\bea}{\begin{eqnarray}}
\newcommand{\eea}{\end{eqnarray}}

\newcommand{\refeq}[1]{Eq.~\eqref{#1}}
\newcommand{\reffig}[1]{Fig.~\ref{#1}}
\newcommand{\reftable}[1]{Table~\ref{#1}}

\newcommand{\iv}[1]{\bm{#1}}

\newcommand{\grad}{\vec{\nabla}}

\newcommand{\mhep}{M_{\text{hep}}}
\DeclareMathOperator{\arctanh}{arctanh}

\graphicspath{{../figs/}{../plots/}{../}}

\preprint{ECT*-10-07}

\begin{document}

\title{Heavy-particle formalism with Foldy-Wouthuysen representation}

\author{Bingwei Long}
\affiliation{European Centre for Theoretical Studies in Nuclear
  Physics and Related Areas (ECT*), I-38123 Villazzano (TN), Italy}
\author{Vadim Lensky}\thanks{On leave from ITEP, Moscow.}
\affiliation{European Centre for Theoretical Studies in Nuclear
  Physics and Related Areas (ECT*), I-38123 Villazzano (TN), Italy}
\affiliation{Theoretical Physics Group, School of Physics and Astronomy,
The University of Manchester, Manchester M13 9PL, United Kingdom}

\date{April 28, 2011}

\begin{abstract}
Utilizing the Foldy-Wouthuysen representation, we use a bottom-up approach to construct heavy-baryon Lagrangian terms,
without employing a relativistic Lagrangian as the starting point.
The couplings obtained this way feature a straightforward $1/m$ expansion, which ensures Lorentz invariance order by order
in effective field theories. We illustrate possible applications with two examples in the context of chiral effective field theory:
the pion-nucleon coupling, which reproduces the results in the literature, and the pion-nucleon-delta coupling, which
does not employ the Rarita-Schwinger field for describing the delta isobar, and hence does not invoke any spurious degrees of freedom.
In particular, we point out that one of the subleading $\pi N \Delta$ couplings used in the literature is, in fact, redundant, and
discuss the implications of this. We also show that this redundant term should be dropped if one wants to use low-energy constants
fitted from $\pi N$ scattering in calculations of $NN\to NN\pi$ reactions.
\end{abstract}

\maketitle

\section{Introduction\label{sec:intro}}

Effective field theories (EFTs) are very useful in describing low-energy physics, in which external momenta $Q$ are much smaller than some high-energy scale $M_\text{hep}$ where the underlying theory kicks in.
The $S$ matrix computed by an EFT is an approximation, namely, an expansion organized in the powers of the small parameter $Q/\mhep$. Effective degrees of freedom (DOFs) of the low-energy theory are not always light particles.
The particles that appear in both the initial and final states (hence, they can not be integrated out)
may have a small momentum $Q$ but a mass $m$ comparable with, or even larger than, $M_\text{hep}$ : $Q \ll M_\text{hep} \lesssim m$. In such cases, one needs to carefully implement these heavy particles
as low-energy effective DOFs so that the ratio $m/M_\text{hep} \gtrsim 1$ will not spoil the EFT expansion. Widely used in many EFTs is heavy-particle formalism \cite{hqet0, mannel92, hbchet}, in which the particles with
$m \sim \mhep$ are allowed only to propagate forward in time, i.e., there are no heavy antiparticle DOFs. In this paper, we consider the application of heavy-particle formalism in chiral effective theory (ChET).

ChET specializes in low-energy interactions among baryons and (pseudo)-Goldstone bosons, which arise due to the fact that chiral symmetry of quantum chromodynamics (QCD) is spontaneously broken.
Since non-Goldstone bosons are all integrated out in ChET, the underlying scale of ChET is set by the mass of the lightest non-Goldstone boson $\sigma$, $m_\sigma \sim 600$ MeV~\cite{Caprini:2005zr}.
Since the nucleon (the lightest baryon) mass, $m_N \simeq 940$ MeV, is not a light scale compared with $m_\sigma$, it is natural to treat baryons with heavy-particle formalism --- heavy-baryon ChET
(HBChET) \cite{hbchet, hbchilag, hbdelta}.

To derive the HBChET Lagrangian, one (probably the most popular) way is nonrelativistic reduction of a relativistic ChET Lagrangian that is built with causal fields (fields that satisfy \emph{microscopic causality}) for baryons, e.g., the Dirac field for the nucleon. Nonrelativistic reduction can be carried out by decoupling low- and high-energy DOFs of the causal baryon fields. In the case of the nucleon, one identifies
the ``large'' and ``small'' components of the Dirac field, respectively, as low- and high-energy DOFs, and then decouples the two sets of DOFs by explicitly integrating out the small components with
the path integral~\cite{mannel92, Bernard:1992qa} or by block diagonalizing the Hamiltonian by the Foldy-Wouthuysen (FW) transformation~\cite{FW50}
(for use of this approach in the context of nucleon-nucleon forces and HBChET, see, e.g., Refs.~\cite{Friar:1999sj,gardestig}).

References ~\cite{mannel92, Bernard:1992qa, FW50, Friar:1999sj, gardestig}, among others, considered only baryon-bilinear operators, exploiting the fact that these operators are quadratic in baryon fields to integrate out or block diagonalize.
It is not immediately clear how a similar method can be applied when four-baryon (or multi-baryon) operators, important for few-nucleon systems, are present. One might wish to treat four-baryon operators as perturbations
to the $NN$ bilinears. However, this is well known not to be the case (see, e.g., Ref.~\cite{nucforce}).

It is not, however, inevitable to rely on the form of a Lagrangian outside the regime of validity of an EFT; only symmetries are what matters. The other approach starts with the nonrelativistic limit, implementing the nucleon and the delta isobar [($\Delta(1232)$)], another important ingredient in ChET~\cite{Pascalutsa:2006up}, as a two- and a four-component spinor, and then enforcing Lorentz invariance order by order with more and more
$1/m_N$ suppressed operators accounted for \cite{galilean, brambilla} (branded differently, reparametrization invariance is a technique in a similar spirit~\cite{rpi}).

In this bottom-up approach, $1/m_N$ expansion of multi-baryon operators is not different 
from that of baryon bilinears (in this connection, see Ref.~\cite{Girlanda:2010ya}, where the construction of all possible
$NNNN$ contact interactions with two derivatives was considered).
The other gain of this approach is that it is convenient to treat the delta isobar \cite{galilean, bwlbira-delta},
because one no longer needs to cope with spurious spin-$1/2$ sectors of the Rarita-Schwinger field, which is commonly used as the causal field for spin-$3/2$ fermions (for a discussion of this and related issues, see
Refs.~\cite{Tang:1996sq,Pascalutsa:1998pw,Krebs:2008zb,Krebs:2009bf} and references therein).

While a nonrelativistic reduction, such as FW transformation, starts from causal relativistic fields and disentangles particle
and antiparticle DOFs, the FW \emph{representation} of the Poincar\'e group~\cite{Foldy56}, where the particle and antiparticle fields
are separated from the beginning, is a technique in the spirit of bottom-up construction (we note that, although credited to the same authors,
the FW \emph{transformation}~\cite{FW50} is not the technique we use in this paper).
By using the FW representation, we present in this paper a systematic machinery to
build HBChET operators that are fixed by Lorentz invariance (and hence suppressed by $1/m_N$), and illustrate the method with
several effective interactions, namely, pion-nucleon and pion-nucleon-delta couplings. With the case of $\pi N N$ coupling being
well known and rather standard, our result for $\pi N\Delta$ coupling is new in the context of HBChET.
More importantly, we find that one of the subleading $\pi N\Delta$ couplings used in the literature is redundant, which
directly affects calculations of many reactions. We discuss the implications of this for $\pi N$ scattering
and the reactions $NN\to NN\pi$.

Our paper is structured as follows. In Sec.~\ref{sec:HBinv}, we revisit the Lorentz invariance of heavy-particle EFT. The FW representation is introduced in Sec.~\ref{sec:FWrep} and its relation with other
Lorentz-covariant fields is discussed in Sec.~\ref{sec:lorcov}. Nucleon-nucleon and nucleon-delta covariant bilinears are discussed in Sec.~\ref{sec:lorcov2}. In Sec.~\ref{sec:piNNpiND}, $\pi N N$ and
$\pi N \Delta$ couplings are used to demonstrate our method of $1/m_N$ expansion, and a discussion of the results is presented.
We summarize and close with a conclusion in Sec.~\ref{sec:summary}.

\section{Lorentz invariance in heavy-particle EFT\label{sec:HBinv}}

When the momentum of heavy particles is much smaller than their mass $m$, Galilean invariance is a good approximation but not a substitute to Lorentz invariance.
As contributions of higher and higher orders are taken into account in order-by-order EFT calculations, the approximation of Lorentz invariance must be improved along the way.
In this section, we review how Lorentz invariance is enjoyed in heavy-particle formalism \cite{hbchet, mannel92}. 
Without a relativistic Lagrangian built with causal baryonic fields as the starting point, microscopic causality will be lost.
Therefore, of particular interest is the following question: How could the bottom-up construction lead to a Lorentz-invariant $S$ matrix?

For definiteness, we consider the $S$ matrix generated by the Dyson series,
\beq
S = T \exp \left[-i\int d^4x\, \mathcal{H}_I(x)
\right] \, , \label{eqn:dyson}
\eeq
where $\mathcal{H}_I(x)$ is the interaction Hamiltonian density in the \emph{interaction picture}.
Here, $T$ indicates, as usual, that the fields are to be time ordered in the expansion.
Starting with a Lagrangian and proceeding with canonical quantization, one does not necessarily end up with an interaction Hamiltonian $\mathcal{H}_I$ that
equals to minus the interaction Lagrangian $-\mathcal{L}_I$ because canonical quantization may produce extra terms \cite{weinbergbook} in some cases.
Nevertheless, for simplicity, we will not concern ourselves with this subtlety and will recklessly assume $\mathcal{H}_I = -\mathcal{L}_I$.  

\begin{figure}
\includegraphics[width=0.8\textwidth]{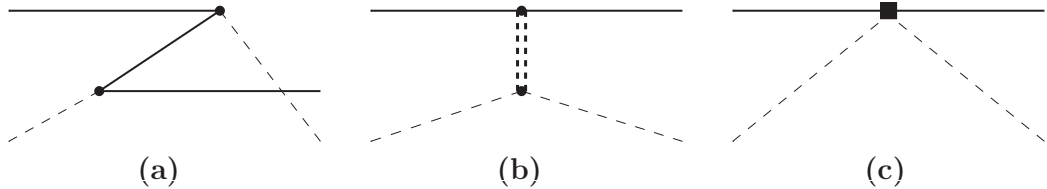}
\caption{\label{fig:shphys}Two types of short-range physics contribute to local EFT operators: (a) intermediate states consisting of a baryon-antibaryon pair and (b) exchanges of a non-Goldstone boson.
Time flows from left to right. Local EFT operators, (c), can be generated by both types of short-range interactions. The solid line represents a baryon, the dashed line a Goldstone boson, and the double-dashed line a non-Goldstone boson.}
\end{figure}

In order for the $S$ matrix to be Lorentz invariant, not only does $\mathcal{L}_I(x)$ need to be invariant, $\mathcal{L}_I(x)$ also needs to be built with causal fields so that $\mathcal{L}_I(x)$
and $\mathcal{L}_I(y)$ will commute with each other when $x-y$ is space-like: microscopic causality. In turn, microscopic causality allows a heavy particle to propagate backward in time.
As a consequence, virtual particle pairs are created and annihilated as intermediate states. Since intermediate states of this sort have energies at least $2m$,
they are integrated out in an EFT and are buried into low-energy constants (LECs). This is exemplified in Fig.~\ref{fig:shphys}(a) with baryon-meson interactions.
Time flows from left to right in the figure, and the baryon internal line propagating backward represents an antibaryon. Having integrated out the baryon-antibaryon pair,
one is left with a local EFT operator as shown by Fig.~\ref{fig:shphys}(c).
Therefore, microscopic causality is preserved order by order in heavy-particle EFT by taking into account the local EFT operators arising from integrating out heavy particle-antiparticle pairs.

In the specific case of HBChET, LECs driven by high-energy intermediate states in Fig.~\ref{fig:shphys}(a) are suppressed by powers of
$1/m_N$, and, in principle, could be computed by explicit integrating out or block diagonalization, as shown in Refs.~\cite{mannel92, Bernard:1992qa, FW50, gardestig}, so that microscopic causality is manifestly satisfied.
Although microscopic causality is crucial for the manifest Lorentz invariance, it is not the sole short-range physics that drives the LECs of HBChET; non-Goldstone bosons propagating could be the other short-range mechanism,
illustrated by Fig.~\ref{fig:shphys} (b).
The contributions by Fig.~\ref{fig:shphys}(b) are suppressed by $1/m_\text{NGB}$, with $m_\text{NGB} \geqslant m_\sigma$ the generic mass of non-Goldstone bosons. Since any local EFT operator generated by Fig.~\ref{fig:shphys}(a)
can as well be generated by Fig.~\ref{fig:shphys}(b), the particular knowledge of the $1/m_N$ contributions does not
improve the predictive power of EFT. Therefore, even if one can compute explicitly the baryon pair-generated contributions to LECs, it
is hardly useful to do so. However, we remark that, in some EFTs, the contributions
to LECs from heavy-particle pairs might dominate. For instance, in heavy-quark effective theory (HQET) for bottom quarks, LECs contributed by electroweak physics are suppressed by the inverse $W$-boson mass,
$1/m_W \ll 1/m_b\,$ \cite{mannel92, rpi}.

To conclude the points we have argued, it is not necessary to constrain the HBChET Lagrangian with microscopic causality.
Lorentz invariance will be enjoyed by the HBChET $S$ matrix as long as $\mathcal{L}_I(x)$ is a Lorentz scalar,
built with the relativistic, isovector pion field $\iv{\pi}$ and forward-propagating baryon fields: a two-component spinor and isospinor $N$ for the nucleon,
and a four-component spinor and isospinor $\Delta$ for the delta isobar.
$\mathcal{L}_I(x)$ is a Lorentz scalar in the sense that
\beq
U_0(\Lambda)\, \mathcal{L}_I\left[\iv{\pi}(x), N(x), \Delta(x) \right]\, U_0^{-1}(\Lambda) = \mathcal{L}_I\left[\iv{\pi} (\Lambda x), N(\Lambda x), \Delta(\Lambda x) \right] \, ,
\label{eqn:lorinvH}
\eeq
where $\Lambda$ is the Lorentz transformation matrix and $U_0(\Lambda)$ is the Lorentz transformation for free $\iv{\pi}$, $N$, and $\Delta$. As pointed out long ago, this level of Lorentz invariance can be achieved without
causal fields~\cite{Foldy56}. 
Note that the fields in the interaction picture satisfy the free equations of motion (EOM), which will be exploited repeatedly in this paper. As we will see, the Lorentz invariance \eqref{eqn:lorinvH} will be enforced
by a set of an infinite number of EFT operators, which are suppressed by inverse powers of $m_N$ and do \emph{not} originate from integrating out intermediate baryon-antibaryon pairs.

\section{Foldy-Wouthuysen representation\label{sec:FWrep}}

A Poincar\'e transformation takes a space-time point $x$ to $x'$,
\beq
x'^{\mu} = \Lambda^{\mu}_{\nu} x^{\nu} + a^{\mu} \, ,
\eeq
with $a^{\mu}$ a four-vector specifying the space-time translation and
$\Lambda^{\mu}_{\nu}$ the Lorentz-transformation matrix.
An infinitesimal Poincar\'e transformation can be written as
\beq
U = 1 - i\vec{\theta}\cdot\vec{J} - i\vec{\xi}\cdot\vec{K} + i\epsilon^\mu P_\mu + \cdots \, ,
\eeq
where $\vec{J}$ are the rotation generators, $\vec{K}$ is the boost, and $P_\mu$ is the space-time translation.
We follow the convention of Ref.~\cite{Foldy56} on the commutation relations among $\vec{J}$, $\vec{K}$, and $P_\mu$,
\begin{align}
[P_i, P_j] & = 0 \, , & [P_i, P_0] & = 0\, , & [J_i, P_j] & = i\epsilon_{ijk} P_k \, , \nonumber \\ 
[J_i, P_0] & = 0 \, , & [J_i, J_j] & = i\epsilon_{ijk}J_k \, , & [P_i, K_j] &= -i \delta_{ij} P_0 \, , \nonumber \\ 
[P_0, K_i] & = -i P_i \, , & [J_i, K_j] & = i\epsilon_{ijk} K_k \, , & [K_i, K_j] &= -i\epsilon_{ijk} J_k \, . \label{eqn:Comm3}
\end{align}

The most commonly used causal fields in building relativistic theories, including the Dirac field,
four-vector, etc., transform under the Poincar\'e group as
\beq
\Phi_{l'}(x) \to \Phi'_{l'}(x) = M(\Lambda)_{l' l}
\Phi_l (x'') \, ,
\label{eqn:Lorrep}
\eeq
where $M(\Lambda)$ are finite-dimension space-time-independent matrices that furnish a (non-unitary) representation of the proper homogeneous Lorentz group, and
\beq
x'' \equiv \Lambda^{-1} (x-a) \, . \label{eqn:xdprime}
\eeq
Transformation~\eqref{eqn:Lorrep} can be symbolically written as 
\beq
\Phi \to M(\Lambda)\Phi \, ,
\label{eqn:Lorrep-simp}
\eeq
with the convention that the left-hand side is evaluated at $x$ while the right-hand side is evaluated at $x''$. Unless pointed out otherwise, the Lorentz transformations in this paper are written
as if the fields were classical.

In the following, we refer to the fields that transform according to \eqref{eqn:Lorrep} as \emph{Lorentz-covariant} fields.
Under an infinitesimal boost, $x''$ and the boosted space-time derivatives are
\beq
t'' = t + \vec{\xi}\cdot\vec{x} \, , \quad \vec{x}\,'' = \vec{x} + \vec{\xi} t  \, ,
\eeq
and
\beq
\partial_t \to \partial_t + \vec{\xi}\cdot\vec{\nabla} \, , \quad \vec{\nabla} \to \vec{\nabla} + \vec{\xi} \partial_t \, . \label{eqn:transd}
\eeq
While being convenient for building relativistic Lagrangians, the Lorentz transformation \eqref{eqn:Lorrep-simp} can not be expanded intuitively in $\partial/m$ because the matrices $M(\Lambda)$ are, by construction, independent of
the momentum or the mass. However, the FW representation allows for such a straightforward expansion~\cite{Foldy56}.

The FW representation of the Poincar\'e group is spanned by the solutions of the relativistic Schr\"odinger equation,
\beq
i\partial_t\chi(x) = \omega\chi(x) \, ,
\label{eqn:relsch}
\eeq
where $\chi(x)$ is a regular SO(3), $(2s+1)$-component spinor with spin $s$ and mass $m$, and $\omega \equiv \sqrt{-\vec{\nabla}^2 + m^2}$. The generators of the FW representation are identified as
\bea
P_\mu &=& \left( \omega,\, i\vec{\nabla} \right) \, , \label{eqn:P0} \\
\vec{J} &=& -i\vec{x}\times\vec{\nabla} + \vec{\Sigma}^{(s)}\, , \label{eqn:J0} \\
\vec{K} &=& \frac{1}{2}\left(\vec{x}\omega + \omega\vec{x}\,\right) + it\vec{\nabla} + \frac{i\vec{\Sigma}^{(s)}\times\vec{\nabla}}{m + \omega} \, , \label{eqn:K0}
\eea
where $\vec{\Sigma}^{(s)}$ are the spin operators for spin-$s$ particles (e.g., $\vec{\Sigma}^{(1/2)}=\vec{\sigma}/2$ with $\vec{\sigma}$ the Pauli matrices), satisfying
\begin{equation}
\left[\Sigma^{(s)}_i , \Sigma^{(s)}_j \right] = i\epsilon_{ijk} \Sigma^{(s)}_k \, .
\end{equation} 
An explicit check shows that the operators defined in Eqs.~\eqref{eqn:P0}--\eqref{eqn:K0}  satisfy the commutation relations of the Poincar\'e algebra \eqref{eqn:Comm3}.

The rotation of fields $\chi$ is standard; therefore, it is routine to build three-scalars (e.g., $\chi^\dagger\chi$), three-vectors (e.g., $\chi^\dagger\vec{\Sigma}^{(s)}\chi$), etc.
The boost is a little more complex,
\bea
\chi(x) \to \chi\,'(x) &=& \left(1 - i\vec{\xi}\cdot\vec{K}\right)\chi(x) \nonumber \\
&=& \left[1 + \frac{i\vec{\xi}\cdot\vec{\nabla}}{2\omega} + \frac{\vec{\xi}\cdot(\vec{\Sigma}^{(s)}\times\vec{\nabla})}{m + \omega} + i\vec{\xi}\cdot\vec{x}\,\left(i\partial_t - \omega\right)\right] \chi(x'') \, .
\label{eqn:transchi}
\eea
Since we are concerned with only the transformation of the free fields, we have dropped
in the boost transformation terms proportional to the free EOM.
The parity and time-reversal transformations of the FW field
$\chi$ are exactly the same as those of a nonrelativistic spinor. In the case of a spin-$1/2$ fermion $\chi_\frac{1}{2}$,
\beq
\chi_\frac{1}{2} \xrightarrow{\mathscr{P}} \pi\, \chi_\frac{1}{2} \, , \quad \chi_\frac{1}{2} \xrightarrow{\mathscr{T}} -i \tau \sigma_2\, \chi^*_\frac{1}{2} \, ,
\eeq
where $\pi$ and $\tau$ are unitary phase factors decided by the species of the particle.

In heavy-particle formalism, it is essential to remove the large phase in $\chi(x)$ by introducing the \emph{heavy-particle field}~\cite{hqet0},
\beq
\Psi(x) \equiv e^{imt} \chi(x) \, ,
\label{eqn:hpf}
\eeq
so that the $\partial/m$ expansion can be facilitated. For example, the EOM for the free field becomes
\beq
i\partial_t\Psi(x) = (\omega - m)\Psi(x) = \left(-\frac{\vec{\nabla}^2}{2m} - \frac{\vec{\nabla}^4}{8 m^3} + \cdots\right) \Psi(x) \, .
\label{eqn:eom}
\eeq
The infinitesimal boost of $\Psi(x)$ is \emph{defined} to be
\bea
\Psi(x) \to \Psi'(x) &\equiv& e^{imt} \chi\,'(x) = e^{imt}\left(1 - i\vec{\xi}\cdot\vec{K}\right)e^{-imt}\Psi(x) \nonumber \\
&=& \left[1 + \frac{i\vec{\xi}\cdot\vec{\nabla}}{2\omega} + \frac{\vec{\xi}\cdot(\vec{\Sigma}^{(s)}\times\vec{\nabla})}{m + \omega} - im\vec{\xi}\cdot\vec{x} \right] \Psi(x'') \, . \label{eqn:transPsi}
\eea
The $-im\vec{\xi}\cdot\vec{x}$ term is important for reproducing the Galilean transformation: the momentum $\vec{p}$ of a nonrelativistic particle shifts to $\vec{p} - m\vec{\xi}$ under the boost, i.e.,
\beq
\Psi^\dagger\grad\Psi \to \Psi^\dagger\grad\Psi -i m\vec{\xi}\,\Psi^\dagger\Psi + \Psi^\dagger \mathcal{O}\left(\xi \frac{\grad^2}{m}\right)\Psi \, .
\eeq
It also serves as a slightly nontrivial reminder that $\Psi$ and $\Psi^\dagger$ must appear in pair in order to have a Galilean invariant operator; hence, the conservation of heavy particle number in EFT.

Suppose that $\chi_\frac{1}{2}$ and $\chi_\frac{3}{2}$ are respectively FW fields with spins $1/2$ and $3/2$, with masses of the nucleon $m_N$ and delta isobar $m_\Delta$.
We introduce heavy-baryon fields for the nucleon and delta isobar, $N$ and $\Delta$, by removing the common nucleon mass $m_N$,
\begin{equation}
N = e^{im_Nt} \chi_\frac{1}{2} \, , \qquad \Delta = e^{im_Nt} \chi_\frac{3}{2} \, .
\label{eqn:defND}
\end{equation} 
So, the free nucleon and delta EOMs are
\begin{align}
i\partial_tN(x) &= \left(-\frac{\vec{\nabla}^2}{2m_N} - \frac{\vec{\nabla}^4}{8 m_N^3} + \cdots\right) N(x) \, , \\
i\partial_t\Delta(x) &= \left(\delta -\frac{\vec{\nabla}^2}{2m_N} + \frac{\delta \vec{\nabla}^2}{2 m_N^3} + \cdots\right) \Delta(x) \, ,
\end{align}
where $\delta$ is the delta-nucleon mass splitting $\delta \equiv m_\Delta - m_N$.

\section{Lorentz-covariant fields\label{sec:lorcov}}

With the boosts of a heavy field $\Psi$ [Eq.~\eqref{eqn:transPsi}], we are already in a position to write down order-by-order Lorentz-invariant operators in terms of $\Psi$.
Consider a spin-$1/2$ heavy-fermion $\Psi$ coupling to a relativistic pseudo-four-vector $A_\mu$. The heavy operator with lowest mass dimension that satisfies parity,
time-reversal, and rotation invariance is $\Psi^\dagger\vec{\sigma}\Psi\cdot\vec{A}$, which transforms under the boost as
\beq
\Psi^\dagger\vec{\sigma}\Psi\cdot\vec{A} \to \Psi^\dagger\vec{\sigma}\Psi\cdot\vec{A} + \Psi^\dagger\vec{\sigma}\Psi \cdot\vec{\xi}\, A_0 + \Psi^\dagger\mathcal{O}\left(\xi A\,\frac{\nabla}{m}\right) \Psi \, .
\eeq
To diminish the Lorentz breaking, one needs a higher-dimension operator $\left(i\Psi^\dagger\vec{\sigma}\cdot\grad\Psi + H.c. \right)A_0$ with a properly tuned coefficient such that the sum of the two has
a Lorentz breaking of higher dimension,
\beq
\Psi^\dagger\vec{\sigma}\Psi\cdot\vec{A} - \frac{1}{2m} \left(i\Psi^\dagger\vec{\sigma}\cdot\grad\Psi + H.c. \right)A_0 \to \text{l.h.s.} + \Psi^\dagger\mathcal{O}\left(\xi A\,\frac{\nabla}{m}\right) \Psi \, .
\eeq
Repeating this procedure, we expect to build a Lorentz-invariant Lagrangian order by order.

The above example suggests that the construction of the effective Lagrangian with the FW fields will be much simplified if one can construct Lorentz-covariant bilinears out of the FW fields, e.g.,
\begin{equation}
a_\mu = \left( - \frac{1}{2m} \left(i\Psi^\dagger\vec{\sigma}\cdot\grad\Psi + H.c. \right) + \cdots, \Psi^\dagger\vec{\sigma}\Psi + \cdots \right) \, .
\end{equation} 
To this end, we would like to establish a field redefinition that maps a FW field onto a Lorentz-covariant field. More precisely, we wish to have a function in terms of the FW field that transforms covariantly but,
of course, does not create the antiparticle (therefore, does not accommodate microscopic causality).
In the case of a spin-$1/2$ field, the function being sought is just the Foldy-Wouthuysen transformation~\cite{FW50},
which can be generalized to particles with arbitrary spin, see, e.g., Ref.~\cite{PhysRev.135.B241}.
We give in the following the results for spin-$1/2$ and spin-$3/2$ fermions, which are of particular relevance in HBChET, in the notation consistent with this paper.
For the detail of the corresponding derivations, the reader is referred to Appendix~\ref{app:details}.
Building Lorentz-covariant fermion bilinears will be discussed in Sec.~\ref{sec:lorcov2}.

The irreducible representations of the homogeneous Lorentz group can be characterized by a pair of integer or half-integer numbers $(A, B)$ (see, e.g., Ref.~\cite{weinbergbook}), with angular momentum
\begin{equation}
j = A + B, A+B-1, \dots, |A-B| \, .
\end{equation}
In our notation, $A$ corresponds to operator $\vec{\mathscr{A}} =\frac{1}{2}\left(\vec{J} - i\vec{K}\right)$ and $B$ to $\vec{\mathscr{B}} = \frac{1}{2}\left(\vec{J} + i\vec{K}\right)$.
Generalizing from the Weyl spinors for $\left(\frac{1}{2}, 0\right)$ and $\left(0, \frac{1}{2}\right)$, we call 
the $(s,0)$ [$(0,s)$] representation a left-handed (right-handed) spinor, denoted by $(2s+1)-$component spinor $\Phi^{(s)}_L$ ($\Phi^{(s)}_R$), which transforms covariantly under boosts as
\beq
\Phi^{(s)}_{L}\to \left[1+\vec\xi\cdot\vec\Sigma^{(s)}\right]\Phi^{(s)}_{L}\,, \qquad
\Phi^{(s)}_{R}\to \left[1-\vec\xi\cdot\vec\Sigma^{(s)}\right]\Phi^{(s)}_{R}\,.
\label{eqn:SSpinors}
\eeq
The idea is to construct, as the building blocks, the $\Phi^{(s)}_L$ and $\Phi^{(s)}_R$ out of the spin-$s$ FW field $\chi_s$, i.e., to construct a certain functional of $\chi_s$ such that
it transforms under boosts according to Eq.~\eqref{eqn:SSpinors}. The field redefinition \eqref{eqn:hpf} will then easily turn the results in terms of the heavy version of the FW fields, $\Psi$.
In order to preserve the rotational properties, $\Phi^{(s)}_L$ or $\Phi^{(s)}_R$ has to take the general form
\beq
\Phi^{(s)}(x)=\sum\limits_{i=0}^{2s}f_i({\vec\nabla}^2)\, T^{(i)}_{j_1,\dots, j_i}\,\nabla_{j_1}\dots\nabla_{j_i}\,\chi_s(x)\equiv F^{(s)}(\vec\nabla)\chi_s(x)\,,
\label{eqn:Transform}
\eeq
where $T^{(i)}_{j_1,\dots, j_i}$ are rank-$i$ tensors built from the generators $\vec\Sigma^{(s)}$ [one can choose them to be irreducible in the sense that $\chi_s^\dagger\,T^{(i)}_{j_1,\dots, j_i}\chi_s$
constitute a spin-$i$ irreducible representation of SO(3)].

\subsection{Spin-$1/2$}

In the spin-$1/2$ case, one needs to look for the left-handed (right-handed) Weyl spinor that transforms under the boost as
\beq
\eta_{L,\,R} \to \left(1\pm \vec\xi\cdot\frac{\vec\sigma}{2}\right)\eta_{L,\,R} \, , \label{eqn:weyltrans}
\eeq
where the left-handed (right-handed) spinor corresponds to the upper (lower) sign.
To remain as a three-spinor and to have desired parity, $\eta_{L,\,R}$ must be related to $\chi_\frac{1}{2}$ as follows,
\beq
\eta_{L,\,R} = \left[f_0(\vec{\nabla}^2) \pm f_1(\vec{\nabla}^2) \vec{\sigma}\cdot\vec\nabla \right] \chi_\frac{1}{2} \, .
\eeq
Applying the above expression and the boost of $\chi_\frac{1}{2}$ [Eq.~\eqref{eqn:transchi}] to the boost of $\eta_{L,\,R}$ [Eq.~\eqref{eqn:weyltrans}], one finds
\beq
\eta_{L,\,R} = \sqrt{\frac{m+\omega}{4\omega}}\left(1 \pm i\frac{\vec\sigma\cdot\vec\nabla}{m+\omega}\right) \chi_\frac{1}{2}\, 
\label{eqn:etalr}
\eeq
(see Appendix~\ref{app:details} for the details).

It is perhaps more conventional to write a Dirac field in terms of $\eta_{L,\, R}$ in the chiral basis where $\gamma^5$ is diagonal,
\beq
\psi_D =
\begin{pmatrix}
\eta_L\\
\eta_R
\end{pmatrix}
= \sqrt{\frac{m+\omega}{4\omega}}
\begin{pmatrix}
\left(1 + i\frac{\vec\sigma\cdot\vec\nabla}{m+\omega}\right) \chi_\frac{1}{2}\\
\left(1 - i\frac{\vec\sigma\cdot\vec\nabla}{m+\omega}\right) \chi_\frac{1}{2}
\end{pmatrix} \label{eqn:diracD}
\, .
\eeq
Here we reproduced the results of Refs.~\cite{FW50, Foldy56}, and the redefinition \eqref{eqn:etalr} is essentially projecting out the large components of the free Dirac field.

\subsection{Spin-$3/2$}

Another case of interest in this paper is the spin-$3/2$ fermion, e.g., the delta isobar.
In this case, the left-handed (right-handed) spinor transforms under the boost as
\beq
\zeta_{L,\,R} \to \left(1\pm \vec\xi\cdot\vec\Sigma\right)\zeta_{L,\,R} \, , \label{eqn:weyltrans32}
\eeq
where the left-handed (right-handed) spinor corresponds to the upper (lower) sign, and we have dropped the superscript $^{\frac{3}{2}}$ in the $4\times 4$ matrices $\vec \Sigma$ that denote the spin operators for
spin-$3/2$ fermion. Again, to keep the correct rotational property and the correct parity, the relation between $\zeta_{L,\,R}$ and the spin-$3/2$ FW field $\chi_\frac{3}{2}$ must be
\beq
\zeta_{L,\,R} = \left[ f_0(\vec{\nabla}^2) \pm f_1(\vec{\nabla}^2) \vec{\Sigma}\cdot\vec\nabla 
+f_2(\vec{\nabla}^2)M_{ij}\nabla_i\nabla_j\pm f_3(\vec{\nabla}^2)T_{ijk}\nabla_i\nabla_j\nabla_k\right] \chi_\frac{3}{2} \, ,
\eeq
where the matrices $M_{ij}$ and $T_{ijk}$ are defined as
\beq\label{eqn:MTdef}
M_{ij}=\frac{1}{2}\left\{\Sigma_i\Sigma_j\right\}-\frac{5}{4}\delta_{ij},\quad T_{ijk}=\frac{1}{6}\left\{\Sigma_i\Sigma_j\Sigma_k\right\}-\frac{41}{60}\left(
\Sigma_i\delta_{jk}+\Sigma_j\delta_{ik}+\Sigma_k\delta_{ij}
\right)\,,
\eeq
where the braces stand for the summation over all the permutations of tensor indices. Note that $M_{ij}$ and $T_{ijk}$ are defined such that
$\chi^\dagger_\frac{3}{2} M_{ij} \chi_\frac{3}{2}$ ($\chi^\dagger_\frac{3}{2} T_{ijk} \chi_\frac{3}{2}$) has only spin-$2$ (spin-$3$) sector
(see Appendix~\ref{app:details} for the details). One finds
\beq
\zeta_{L,\,R} = \sqrt{\frac{m+\omega}{4\omega}} \left[\frac{\omega}{m} \pm 
i\frac{(6\omega+4m)\vec\Sigma\cdot\vec\nabla}{5m(m+\omega)}-\frac{M_{ij}\nabla_i\nabla_j}{m(m+\omega)}
\mp i\frac{2T_{ijk}\nabla_i\nabla_j\nabla_k}{3m(m+\omega)^2}\right] \chi_\frac{3}{2} \, . \label{eqn:zetalr}
\eeq
Most importantly, one no longer needs to deal with any spurious DOFs because, unlike in the case of Rarita-Schwinger field, there is only one spin-$3/2$ sector and no other spin-$1/2$ sector.

\section{Lorentz-covariant bilinears\label{sec:lorcov2}}

The bilinears sought after are, in general, tensors of integer rank $n$, which are direct products of $n$ vectors, which can, in turn, be decomposed
into irreducible terms $(A, B)$ with $A = n/2, n/2-1, \dots$ and $B = n/2, n/2-1, \dots$, by various symmetrizations, antisymmetrizations and extracting traces.
For example, $(0, 0)$ is a scalar, $(1,0)\oplus(0,1)$ an antisymmetric rank-2 tensor, and $(1, 1)$ a symmetric rank-2 tensor. A vector (or a pseudovector) is represented by $\left(\frac{1}{2}, \frac{1}{2}\right)$.

To accommodate parity, one needs to use the direct sum of $\Phi^{(s)}_L$ and $\Phi^{(s)}_R$ in building fermion bilinears: $(s, 0)\oplus(0, s)$. Since there are no antiparticle DOFs, $\Phi^{(s)}_L$ and $\Phi^{(s)}_R$
are not independent of each other. 

\subsection{$NN$ bilinears}
Because
\begin{equation}
\left[\left(\frac{1}{2}, 0\right)\oplus\left(0, \frac{1}{2}\right)\right] \otimes \left[\left(\frac{1}{2}, 0\right)\oplus\left(0, \frac{1}{2}\right)\right] 
= (0, 0) \oplus (0, 0) \oplus \left(\frac{1}{2}, \frac{1}{2}\right) \oplus \left(\frac{1}{2}, \frac{1}{2}\right) \oplus \left[(1, 0)\oplus(0, 1)\right] \, ,
\end{equation} 
nucleon-nucleon bilinears include a scalar ($s$), a pseudoscalar ($p$), a vector ($v^\mu$), a pseudovector ($a^\mu$) and an antisymmetric tensor ($F^{\mu\nu}$), as is well known.
Our procedure of assembling these bilinears is as follows. Consider at first the boosts of, for instance, the following three-scalars and three-vectors,
\bea
\eta_L^\dagger\eta_L\to\eta_L^\dagger\eta_L-\vec\xi\cdot(-\eta^\dagger_L\vec\sigma\eta_L),&\quad&\eta^\dagger_L\vec\sigma\eta_L\to\eta^\dagger_L\vec\sigma\eta_L -\vec\xi(-\eta_L^\dagger\eta_L)\,,\label{eqn:lorentzboosts1}\\
\eta_R^\dagger\eta_R\to\eta_R^\dagger\eta_R-\vec\xi\cdot(\eta^\dagger_R\vec\sigma\eta_R),&\quad&\eta^\dagger_R\vec\sigma\eta_R\to\eta^\dagger_R\vec\sigma\eta_R -\vec\xi(\eta_R^\dagger\eta_R)\,.\label{eqn:lorentzboosts2}
\eea
Note also that $\eta_R$ and $\eta_L$ get interchanged under spatial reflections. This allows one to conclude that $(\eta_L^\dagger\eta_L\pm\eta_R^\dagger\eta_R,\pm\eta^\dagger_R\vec\sigma\eta_R-\eta^\dagger_L\vec\sigma\eta_L)$
is a contravariant four-vector (pseudovector). Analogous considerations can be applied to all other SO(3) bilinears. Listed in Table \ref{tbl:bihalf} are all of the $NN$ covariant bilinears.

\begin{table}
\caption{\label{tbl:bihalf} Lorentz-covariant bilinears built out of the spin-$1/2$ FW field, where $\eta_{L,\,R}$ are defined in Eq.~\eqref{eqn:etalr}.
}
\begin{tabular}{|c|c|}
\hline
$s$ & $s=\eta_L^\dagger\eta_R+\eta_R^\dagger\eta_L$ \\
\hline
$p$ & $p=\eta_L^\dagger\eta_R-\eta_R^\dagger\eta_L$\\
\hline
$v^\mu$ & $v^\mu=\left(\eta_L^\dagger\eta_L+\eta_R^\dagger\eta_R,\eta_R^\dagger\vec\sigma\eta_R-\eta_L^\dagger\vec\sigma\eta_L\right)$\\
\hline
$a^\mu$ & $a^\mu=\left(\eta_L^\dagger\eta_L-\eta_R^\dagger\eta_R,-\eta_R^\dagger\vec\sigma\eta_R-\eta_L^\dagger\vec\sigma\eta_L\right)$\\
\hline
$F^{\mu\nu}$ & $F^{0i}=i\eta_L^\dagger\sigma_i\eta_R-i\eta_R^\dagger\sigma_i\eta_L,\ F^{ij}=\epsilon_{ijk}\left(\eta_L^\dagger\sigma_k\eta_R+\eta_R^\dagger\sigma_k\eta_L\right)$\\
\hline
\end{tabular}
\end{table}

\subsection{$N\Delta$ bilinears}
The product of a spin-$1/2$ and a spin-$3/2$ fermion is decomposed as
\begin{equation}
\begin{split}
& \left[\left(\frac{1}{2}, 0\right)\oplus\left(0, \frac{1}{2}\right)\right] \otimes \left[\left(\frac{3}{2}, 0\right)\oplus\left(0, \frac{3}{2}\right)\right] \\
& = \left[(1, 0) \oplus (0, 1)\right] \oplus \left[\left(\frac{3}{2}, \frac{1}{2}\right) \oplus \left(\frac{1}{2}, \frac{3}{2}\right)\right] \oplus \left[(2, 0)\oplus(0, 2)\right] \, ,
\end{split}
\label{eqn:NDdecom}
\end{equation} 
where $(1, 0) \oplus (0, 1)$ is a rank-2 antisymmetric tensor ($G^{\mu \nu}$), $\left(\frac{3}{2}, \frac{1}{2}\right) \oplus \left(\frac{1}{2}, \frac{3}{2}\right)$ and $(2, 0)\oplus(0, 2)$ a rank-3 ($F^{\mu \nu \lambda}$)
and a rank-4 tensor ($H^{\mu \nu \lambda \rho}$), respectively, with the following symmetry properties,
\beq
F^{\mu\nu\lambda}=-F^{\mu\lambda\nu},\quad H^{\mu\nu\lambda\rho}=H^{\lambda\rho\mu\nu}=-H^{\nu\mu\lambda\rho}=-H^{\mu\nu\rho\lambda}\,.
\label{eqn:tensorsym}
\eeq

The SO(3) bilinears at our disposal are
\beq
\eta_{L,R}^\dagger\vec S\zeta_{L,R}, \quad \eta_{L,R}^\dagger \sigma_i S_j\zeta_{L,R}\,,
\eeq
where $S_i$ are the $2\times 4$
transition matrices in spin space, normalized so that
\beq
S_i {S_j}^{\dagger} = \frac{1}{3} \left(2 \delta_{ij} - i
\epsilon_{ijk} \sigma_k \right)\, .
\label{eqn:normS}
\eeq
They have the property
\beq
\sigma_i S_j - \sigma_j S_i = -\frac{\sigma_i}{2} S_j + S_j
\Sigma_i = -i\epsilon_{ijk} S_k \, .
\label{eqn:sSalgebra}
\eeq
Using \eqref{eqn:weyltrans32} and \eqref{eqn:sSalgebra}, one can get the boost rules for the bilinears in full analogy to \eqref{eqn:lorentzboosts1} and \eqref{eqn:lorentzboosts2}, and ultimately build the tensors.
For instance, one gets
\bea
\eta_{L}\vec S\zeta_{R}\to\eta_{L}\vec S\zeta_{R}-i\eta_{L}(\vec\xi\times\vec S)\zeta_{R},\quad \eta_{R}\vec S\zeta_{L}\to \eta_{R}\vec S\zeta_{L}+i\eta_{R}(\vec\xi\times\vec S)\zeta_{L}\,,
\eea
which allows one to conclude that $\eta_{L} S_i\zeta_{R}+\eta_{R} S_i\zeta_{L}$ and $-i\epsilon_{ijk}(\eta_{L} S_k\zeta_{R}-\eta_{R} S_k\zeta_{L})$ transform, respectively, as $(0i)$ and $(ij)$
components of an antisymmetric tensor $G^{\mu\nu}=-G^{\nu\mu}$.
Performing analogous calculations for the remaining bilinears, one
obtains the explicit expressions for these tensors, as given in Table~\ref{tbl:bi3half}.

Of practical use for EFT calculations are the first few $1/m$ terms of Lorentz-covariant bilinears. In Appendix~\ref{app:exp} we give the expansion of bilinears defined in Tables~\ref{tbl:bihalf} and \ref{tbl:bi3half},
up to and including $\mathcal{O}\left[(\nabla/m)^2\right]$ terms.

\begin{table}
\caption{\label{tbl:bi3half}Lorentz-covariant bilinears built out of the spin-$1/2$ and the spin-$3/2$ FW fields, where $\eta_{L,\,R}$ and $\zeta_{L,\,R}$
are defined in Eqs.~\eqref{eqn:etalr} and \eqref{eqn:zetalr}, respectively.
$\Omega_{ij}$ is defined as $\Omega_{ij}\equiv (\sigma_i S_j+\sigma_j S_i)/2$.}
\begin{tabular}{|c|p{0.7\textwidth}|}
\hline
$G^{\mu\nu}$ & $G^{0i}=\eta_L^\dagger S_i\zeta_R+\eta_R^\dagger S_i\zeta_L,\ G^{ij}=-i\epsilon_{ijk}(\eta_L^\dagger S_k\zeta_R-\eta_R^\dagger S_k\zeta_L)$ \\
\hline
$F^{\mu\nu\lambda}$ & $F^{00i}=\eta_L^\dagger S_i\zeta_L+\eta_R^\dagger S_i\zeta_R,\ F^{0ij}=i\epsilon_{ijk}(\eta_L^\dagger S_k\zeta_L-\eta_R^\dagger S_k\zeta_R),$\par
$F^{ij0}=\eta_L^\dagger \sigma_i S_j\zeta_L-\eta_R^\dagger \sigma_i S_j\zeta_R,\ F^{ijk}=-i\epsilon_{jkl}(\eta_L^\dagger \sigma_i S_l\zeta_L+\eta_R^\dagger \sigma_i S_l\zeta_R)$
\\
\hline
$H^{\mu\nu\lambda\rho}$ & $H^{0i0j}=\eta_L^\dagger \Omega_{ij}\zeta_R-\eta_R^\dagger \Omega_{ij}\zeta_L,\ H^{0ijk}=-i\epsilon_{jkl}(\eta_L^\dagger \Omega_{il}\zeta_R+\eta_R^\dagger \Omega_{il}\zeta_L),$\par
$H^{ijkl}=-\epsilon_{ijm}\epsilon_{kln}(\eta_L^\dagger \Omega_{mn}\zeta_R-\eta_R^\dagger \Omega_{mn}\zeta_L)$
\\
\hline
\end{tabular}
\end{table}

\section{$\pi NN$ and $\pi N \Delta$ couplings\label{sec:piNNpiND}}
As applications of our approach to HBChET, we will consider couplings of a spin-$1/2$ field ($N$) to the gradient of the pion field ($\partial_\mu\pi^a$ with $a$ the isospin index) and
the transition of $N$ to a spin-$3/2$ field ($\Delta$) via emitting a pion.

With the pseudovector $NN$ bilinear [Eq.~\eqref{eqn:NNa}], the $NN$ axial-vector coupling and the first terms of its $1/m_N$ expansion are
\begin{equation}
\begin{split}
\mathcal{L}_{\pi NN} & =-g_A{\overline{\psi}}_D\tau^a \gamma^5 \gamma^\mu\psi_D\, \frac{\partial_\mu\pi^a}{2f_\pi} \\
& = g_A N^\dagger \tau^a \vec\sigma N \cdot \frac{\vec\nabla\pi^a}{2f_\pi}
-\frac{g_A}{2m_N}\left[iN^\dagger\tau^a\vec\sigma\cdot\vec{\nabla} N+H.c.\right]\frac{\dot{\pi^a}}{2f_\pi} \\
& \quad {}+\frac{g_A}{4m_N^2}\left[N^\dagger\tau^a\vec\sigma \vec{\nabla}^2 N+(\vec{\nabla}N)^\dagger\tau^a(\vec\sigma\cdot\vec{\nabla})N + H.c.\right]\cdot\frac{\vec\nabla\pi^a}{2f_\pi} + \cdots \, ,
\label{eqn:LpiNN} 
\end{split}
\end{equation} 
where $\psi_D$ is defined in Eq.~\eqref{eqn:diracD} and $f_\pi \simeq 93$ MeV is the pion decay constant.. This expansion coincides with the well-known result in, e.g., Ref.~\cite{hbchilag}.

As seen in Eq.~\eqref{eqn:NDdecom}, there is no pseudovector $N\Delta$ bilinear. The way out is to 
invoke a contraction of tensor bilinears with the derivatives of $N$ or $\Delta$.
Consider a coupling of $\partial_\mu \pi$ to $G^{\mu\nu}$.
We denote by $\partial_\mu^N$ ($\partial_\mu^\Delta$) the derivative that acts on the nucleon (delta) field.
Since the following equality holds:
\beq
\partial_\mu^\Delta G^{\mu\nu}\partial_\nu\pi=\partial_\mu(G^{\mu\nu}\partial_\nu\pi)-\partial^N_\mu G^{\mu\nu}\partial_\nu\pi-G^{\mu\nu}\partial_\mu\partial_\nu\pi\,,
\eeq
where the last term in the r.h.s.\ vanishes due to the antisymmetricity of $G^{\mu\nu}$, and the first term therein is a total derivative, the only independent $\pi N\Delta$ coupling that one can construct with $G^{\mu\nu}$
is
\bea
\mathcal{L}_{\pi N\Delta}&=& \frac{h_A}{m_\Delta} \left[i\partial_\mu^\Delta G^{\mu\nu}+H.c.\right]\, \frac{\partial_\nu\pi}{2f_\pi}\,,
\label{eqn:LpiND1}
\eea
where we have suppressed for the moment the isospin index, and pseudovector $P^\nu \equiv \left[i\partial_\mu^\Delta G^{\mu\nu}+H.c.\right]$ has the structure
\bea
P^{0} &=& -i\left[\eta_L^\dagger\, (\vec S\cdot\vec\nabla)\, \zeta_R+\eta_R^\dagger\, (\vec S\cdot\vec\nabla)\,\zeta_L\right]+H.c.\,,\\
\vec P &=& i\left( \eta_L^\dagger\, \vec S\, \partial_t\zeta_R + \eta_R^\dagger\, \vec S\, \partial_t\zeta_L\right) - \left[\eta_L(\vec\nabla\times\vec S)\zeta_R-\eta_R(\vec\nabla\times\vec S)\zeta_L\right] + H.c.\,.
\eea
The $1/m_N$ expansion of \eqref{eqn:LpiND1} is straightforward as long as one switches to heavy fields and uses the EOMs for $\Delta$ to get rid of their time derivatives,
\begin{equation}
\begin{split}
\mathcal{L}_{\pi N \Delta} &= h_A \left[N^\dagger \vec S T^a\Delta+H.c.\right]\cdot\frac{\vec\nabla\pi^a}{2f_\pi}
-\frac{h_A}{m_N}\left[i N^\dagger \vec S\cdot \vec{\nabla} T^a\Delta +H.c.\right]\frac{\dot{\pi^a}}{2f_\pi} \\ 
&+\delta\frac{h_A}{m_N^2}\left[i N^{\dagger} \vec{S}\cdot\vec{\nabla}T^a\Delta + H.c.\right]\frac{\dot{\pi^a}}{2f_\pi}
+\frac{h_A}{2 m_N^2} 
\left[\left(N^{\dagger} \vec{S}\vec{\nabla}^2 T^a\Delta - N^{\dagger} (\vec{S}\cdot\vec{\nabla})\vec{\nabla}T^a\Delta\right)
+H.c.\right] \cdot\frac{\vec\nabla\pi^a}{2f_\pi}\\
&+ \frac{h_A}{8m_N^2}\left[\left(\delta_{lm} N^{\dagger} \vec{S}\cdot\vec{\nabla}T^a\Delta 
 +3  N^{\dagger} S_l\nabla_m \Delta
 -2i \epsilon_{ijl}N^{\dagger}\Omega_{im} \nabla_jT^a\Delta  \right) + H.c.\right] \frac{\nabla_l \nabla_m\pi^a}{2f_\pi} + \cdots \,,
\label{eqn:LpiND1_exp}
\end{split}
\end{equation} 
where $T^a$ are the isospin analogs of $S^a$, normalized as in Eq.~\eqref{eqn:normS}. Here
the first two terms in the expansion give the well-known non-relativistic result~\cite{galilean}.

Further couplings can be built by a contraction of $F^{\mu\nu\lambda}$ with two derivatives. One can choose to work with $\partial_\mu^N\partial_\nu^\Delta F^{\mu\nu\lambda}\partial_\lambda\pi$
and $\partial_\nu^N\partial_\lambda^\Delta F^{\mu\nu\lambda}\partial_\mu\pi$, with other possibilities dependent on these two via the symmetry property [Eq.~\eqref{eqn:tensorsym}] by partial integrations.
However, with the help of the EOM for the nucleonic Dirac spinor $\psi_D$, one of these two couplings, $\partial_\mu^N\partial_\nu^\Delta F^{\mu\nu\lambda}\partial_\lambda\pi$, can be shown to be equivalent 
to $\partial_\mu^\Delta G^{\mu\nu}\partial_\nu\pi$, thus leaving us with only one independent term, $\partial_\nu^N\partial_\lambda^\Delta F^{\mu\nu\lambda}\partial_\mu\pi$,
\bea
\mathcal{L}_{\pi N\Delta}'&=& \frac{b}{m_N m_\Delta}\left[\partial_\nu^N\partial_\lambda^\Delta F^{\mu\nu\lambda}+H.c.\right]\, \frac{\partial_\mu\pi}{2f_\pi}\,,
\label{eqn:LpiND2}
\eea
where $b$ is the corresponding dimensionless coupling constant.
The first two terms of the expansion of $\mathcal{L}'_{\pi N\Delta}$ in powers of $\nabla/m_N$
can be shown to be equal to the first two terms of the corresponding expansion of $\mathcal{L}_{\pi N\Delta}$, times the small factor $\delta/m_N$:
\bea
\mathcal{L}_{\pi N\Delta}'&=& b\,\frac{\delta}{m_N}\left\{\left[N^\dagger\vec S T^a\Delta+H.c.\right]\cdot\frac{\vec{\nabla}\pi^a}{2f_\pi}
-\frac{1}{m_N}\left[iN^\dagger\vec S\cdot\vec{\nabla} T^a \Delta+H.c.\right]\frac{\dot{\pi}^a}{2f_\pi}\right\}+\cdots\,.
\eea
Therefore, unless $\delta$ is a variable that could depend on the number of colors in QCD,
$\mathcal{L}'_{\pi N\Delta}$ is equivalent to $\mathcal{L}_{\pi N\Delta}$, up to and including $\mathcal{O}(p^3)$, where $p$ stands generically for a small momentum factor such as $\nabla$, $\delta$, etc.

Finally, considering a contraction of $H^{\mu\nu\lambda\rho}$ with three derivatives and taking into account the symmetry of $H^{\mu\nu\lambda\rho}$,
one arrives at the only independent coupling, $\partial_\mu^\Delta\partial_\nu^N\partial_\lambda^\Delta H^{\mu\nu\lambda\rho}\partial_\rho\pi$,
\bea
\mathcal{L}_{\pi N\Delta}''&=& \frac{d}{m_N m_\Delta^2}\left[i\partial_\mu^\Delta\partial_\nu^N\partial_\lambda^\Delta H^{\mu\nu\lambda\rho}+H.c.\right]\, \frac{\partial_\rho\pi}{2f_\pi}\,.
\label{eqn:LpiND3}
\eea
The expansion of this Lagrangian in powers of $\nabla/m$ starts at $\mathcal{O}\left[(\nabla/m)^2\right]$, which can be proved with the help of integration by parts:
\bea
\mathcal{L}_{\pi N\Delta}''&=&- \frac{d}{2m_N^2}\left[N^\dagger\vec S T^a\Delta+H.c.\right]\cdot\frac{\vec{\nabla}(\nabla^2\pi^a)}{2f_\pi}+\cdots\,.
\eea
Using EOMs for pion, $N$, and $\Delta$ fields, one can show that $\mathcal{L}_{\pi N\Delta}''$, in fact, starts introducing
new $\pi N\Delta$ operators at least at $\mathcal{O}(p^4)$, similarly to $\mathcal{L}_{\pi N\Delta}'$.

Therefore, we conclude that the couplings \eqref{eqn:LpiND2} and \eqref{eqn:LpiND3} start producing new operators
at as early as $\mathcal{O}(p^4)$, and thus the only independent $\pi N \Delta$ vertex
up to $\mathcal{O}(p^3)$ is \eqref{eqn:LpiND1}, expanded out in Eq.~\eqref{eqn:LpiND1_exp}. This is, however, at odds with some results in the literature.
References \cite{hbdelta,Hemmert:1997ye,threshold-delta2} consider an additional $\pi N \Delta$ coupling with an undetermined LEC,
$b_3 + b_8 \sim 1/M_\text{hep}$, the leading term in the $1/m_N$ expansion of which is an $\mathcal{O}(p^2)$ operator,
\beq
\mathcal{L}_{\pi N \Delta}^{b_3+b_8} = -2(b_3+b_8)\left(i N^\dagger\vec{S}T^a\Delta +  H.c.\right)
\cdot \frac{\vec{\nabla}\dot\pi^a}{2f_\pi}\,,
\label{eqn:b3b8}
\eeq
If our conclusion is correct, one must be able to show that $b_3+b_8$ is redundant.

To show this more explicitly for its leading term, $\mathcal{L}_{\pi N \Delta}^{b_3+b_8}$, at the level of heavy-baryon operators, we first write down the leading, $\mathcal O(p)$, HBChET Lagrangian that has both baryon EOMs,
the leading $\pi N N$, $\pi N \Delta$ and $\pi \Delta \Delta$ couplings,
\begin{equation}
\begin{split}
\mathcal{L}^{(0)} & = iN^\dagger \partial_t N + g_A N^\dagger \tau^a \vec{\sigma} N \cdot \frac{\vec\nabla\pi^a}{2f_\pi} \\
& \quad {} + \Delta^\dagger (i\partial_t - \delta) \Delta + 4g^{\Delta}_A \Delta^\dagger \,t_{(\frac{3}{2})}^a \vec{\Sigma} \Delta \cdot \frac{\vec\nabla\pi^a}{2f_\pi} + h_A \left( N^\dagger \vec S T^{a}
\Delta + H.c.\right) \cdot \frac{\vec\nabla\pi^a}{2f_\pi} + \{\Psi^\dagger\Psi\Psi^\dagger\Psi\} \, ,
\label{eqn:Lag0}
\end{split}
\end{equation}
where $g^{\Delta}_A$ is the $\Delta$ axial coupling constant, $t_{(\frac{3}{2})}^a$ are the isospin $3/2$ generators, $\Psi$ is generic baryon field and $\{\Psi^\dagger\Psi\Psi^\dagger\Psi\}$ are non-derivative
four-baryon operators whose details are irrelevant.

Using partial integration, we can choose the independent $\pi N \Delta$ operators that have a time \emph{and} 
a spatial derivative, hence of $\mathcal O(p^2)$, to be
\begin{equation}
h_1\left(i\dot{N}^\dagger\vec{S}T^a\Delta + H.c. \right)\cdot\frac{\grad\pi^a}{2f_\pi} + h_2\left(iN^\dagger\vec{S}T^a\dot{\Delta} + H.c. \right)\cdot\frac{\grad\pi^a}{2f_\pi} \, ,
\end{equation} 
where $h_{1, 2} \sim 1/M_\text{hep}$. This absorbs all the effects of Eq.~\eqref{eqn:b3b8}. We note that, in order to respect chiral symmetry, there has to be at least one derivative on $\pi$.
We now want to get rid of interactions that have baryon time derivatives, as commonly practiced in heavy-baryon EFTs.
The following field redefinition,
\begin{align}
N &\to N + h_1 T^a\vec{S}\cdot\frac{\grad\pi^a}{2f_\pi}\Delta \, , \\
\Delta &\to \Delta - h_2 {T^a}^\dagger\vec{S}^\dagger\cdot\frac{\grad\pi^a}{2f_\pi}N \, ,
\end{align}
generates a $\mathcal O(p^2)$ variation,
\begin{equation}
\begin{split}
\mathcal L \to \mathcal{L} &- h_1\left(i\dot{N}^\dagger\vec{S}T^a\Delta + H.c. \right)\cdot\frac{\grad\pi^a}{2f_\pi} - h_2\left(iN^\dagger\vec{S}T^a\dot{\Delta} + H.c. \right)\cdot\frac{\grad\pi^a}{2f_\pi} \\
&\quad + \delta\, h_2 \left(N^\dagger\vec S T^{a}\Delta + H.c.\right)\cdot\frac{\vec\nabla\pi^a}{2f_\pi} + \mathcal{O}(p^2)\{\pi\pi\Psi^\dagger\Psi\} + \mathcal{O}(p^2)\{\pi\Psi^\dagger\Psi\Psi^\dagger\Psi\} + \cdots \, ,
\end{split}
\end{equation} 
where the ellipsis includes $\mathcal O(p^3)$ operators.
The first line shows that the field redefinition was designed to eliminate both $h_1$ and $h_2$ terms, at the expense of generating other operators
that are listed in the second line. It is important that (\textit{i}) these generated operators
do not have baryonic time derivatives and they can be absorbed into the existing $\mathcal O(p^2)$
HBChET Lagrangian, and that (\textit{ii}) the $b_3 + b_8$ operator, absorbed earlier into $h_1$ and $h_2$ operators by partial integration, is not reincarnated.
The first operator in the second line can be lumped into the $h_A$ term by redefining $h_A$: $h_A + \delta h_2 \to h_A$.
Alternatively, the rearrangement of the Lagrangian evoked by the field redefinition can also be
realized by applying EOMs of the nucleon and delta fields.
After all these operations, the effects of the $b_3 + b_8$ term are transformed into various terms,
\begin{equation}
\begin{split}
\mathcal{L}_{\pi N \Delta}^{b_3+b_8} &\to 2(b_3+b_8)\left[
\delta\,\left( N^\dagger T^a\vec{S}\Delta + H.c.\right)\cdot \frac{\vec\nabla\pi^a}{2f_\pi}
- \frac{8}{9} h_A N^\dagger N \left(\frac{\vec\nabla\pi^a}{2f_\pi}\right)^2 \right. \\
& \left. \quad {}+ \frac{2}{9} h_A N^\dagger\epsilon_{ijk}\epsilon^{abc}\sigma_k\tau^cN \frac{\nabla_i\pi^a}{2f_\pi}\frac{\nabla_j\pi^b}{2f_\pi}
\vphantom{\left(\frac{\vec\nabla\pi^a}{2f_\pi}\right)^2}
+\cdots \right]\,.
\label{eqn:b3b8trans}
\end{split}
\end{equation}
Here we only explicitly listed the $\pi N \Delta$ and $\pi \pi NN$ pieces that will be useful later; subsumed in the ellipsis are such terms as $\pi \pi N \Delta$, $\pi \pi \Delta \Delta$, etc.

As promised, we showed that $b_3 + b_8$ is not an independent parameter.
To further illustrate this point in practice, we show in Appendix \ref{app:piN} that, in Ref.~\cite{Krebs:2007rh},
while two fits of LECs to low-energy $\pi N$ scattering data yield two different sets of LECs, this apparent difference is purely due to the arbitrary choice of a set
containing a redundant parameter. We also discuss possible implications of this for other reactions, in particular, $NN\to NN\pi$.

\section{Conclusion\label{sec:summary}}

We demonstrated how one can build heavy-particle Lorentz-invariant Lagrangians with fields that furnish the FW representation for the Poincar\'e group. At the core of the method are Eqs.~\eqref{eqn:etalr}
and \eqref{eqn:zetalr}, which map the FW fields onto the more conventional, Lorentz-covariant left- and right-handed spinors, in the cases of spin-$1/2$ and -$3/2$ corresponding to the nucleon and delta isobar in HBChET.
We also built covariant $N N$ and $N \Delta$ bilinears (Tables~\ref{tbl:bihalf} and~\ref{tbl:bi3half}) and their $1/m_N$ expansions (Appendix~\ref{app:exp}) that are useful in HBChET.
A Lorentz-invariant interaction can thereby be assembled with the FW fields and, in the mean time, can be easily expanded in powers of ${\nabla}/m$.

The machinery we presented here provides a couple of advantages in working out $1/m$ expansion over the explicit integrating out or block diagonalization used in, e.g., Refs.~\cite{mannel92, Bernard:1992qa, FW50, gardestig}.
Firstly, it is natural to apply the same method to multi-fermion operators such as $NNNN$, $NNN\Delta$, etc., whereas it is more difficult to do so in integrating out or block diagonalization because the Lagrangian 
is no longer quadratic in baryon fields. Secondly, when treating the delta or other high-spin baryons, one no longer needs to deal with spurious DOFs.

We illustrated the technique with the examples of $\pi NN$ and $\pi N \Delta$ couplings. The well-known $1/m_N$ expansion of $\pi N N$ is explicitly reproduced up to $\mathcal{O}(p^3)$. $\pi N \Delta$
is more interesting because there is no $N \Delta$ bilinear that is a four-pseudovector. We analyzed all possible $\pi N\Delta$ couplings up to $\mathcal{O}(p^3)$
and found that there is only one independent $\pi N \Delta$ coupling up to and including $\mathcal{O}(p^3)$.
In Appendix~\ref{app:piN}, we use low-energy $\pi N$ scattering to further illustrate that the employment in the literature of the $\mathcal{O}(p^2)$ $\pi N \Delta$ operator, with LEC $b_3 + b_8$, is redundant.
We also discuss there the possible implications of this redundancy for calculations of the reactions $NN\to NN\pi$. In particular, we show that the inclusion of $b_3+b_8$ at $\mathcal{O}(p^2)$ in $\pi N$ scattering
can lead to unnaturally large variations of some terms in $NN\to NN\pi$, which can be cured by demoting the $b_3+b_8$ term to $\mathcal{O}(p^4)$ at the level of the HBChET Lagrangian.

\acknowledgments
We thank U.~van Kolck, M.~Birse, J.~McGovern, J.~Friar and V.~Pascalutsa for valuable discussions on this subject and comments on the manuscript, and T.~Goldman for bringing to our attention J.~Friar's early work.
BwL is grateful for hospitality to the Helmholtz Institut f\"ur Strahlen und Kernphysik at the University of Bonn, where part of this work was done. Some of the calculations in this work were performed with the help of symbolic
computation program FORM~\cite{FORM}. Figures were drawn with the help of package JaxoDraw~\cite{JaxoDraw}. 

\appendix
\section{The redundancy of $\mathcal{L}_{\pi N\Delta}^{b_3+b_8}$ at $\mathcal{O}(p^2)$ and the reactions $NN\to NN\pi$\label{app:piN}}

Here, we examine in more detail whether the $\pi N \Delta$ LEC $b_3 + b_8$ plays any role in a specific process, $\pi N$ scattering below the delta threshold,
and discuss the significance of $\mathcal{L}_{\pi N \Delta}^{b_3+b_8}$ being redundant for calculations of other processes, in particular, the reactions $NN\to NN\pi$.

Up to the next-to-leading order (NLO) in the so-called small-scale expansion (SSE)~\cite{Hemmert:1997ye,threshold-delta2}, the diagrams contributing
to $\pi N$ scattering are all trees and were discussed in detail, e.g., in Refs.~\cite{threshold-delta2, Krebs:2007rh}. Aside from the leading Lagrangian \eqref{eqn:Lag0}, one also needs
$\mathcal{O}(p^2)$ seagull terms~\cite{hbchilag},
\beq
\mathcal{L}^{(1)}_{\pi\pi NN}=N^\dagger\left[4(c_2+c_3)\left(\frac{\dot\pi^a}{2f_\pi}\right)^2-4c_3\left(\frac{\vec\nabla\pi^a}{2f_\pi}\right)^2
-2c_4\,\epsilon_{ijk}\sigma_k\epsilon^{abc}\tau^c\frac{\nabla_i\pi^a}{2f_\pi}\frac{\nabla_j\pi^b}{2f_\pi}\right]N\,,
\label{eqn:pipinn}
\eeq
where we have suppressed $1/m_N$ corrections to $\pi\pi NN$ LECs $c_2$, $c_3$, and $c_4$. Comparing the above Lagrangian and Eq.~\eqref{eqn:b3b8trans}, one can redefine $h_A$ and $c_i$'s
to eliminate $b_3 + b_8$ from the $\pi N\Delta$ Lagrangian at this order, $\mathcal{O}(p^2)$:
\begin{equation}
\begin{split}
\bar h_A&=h_A+2\delta(b_3+b_8)\,,\  \bar c_2=c_2-\frac{4}{9}h_A(b_3+b_8)\,,\\
\bar c_3&=c_3+\frac{4}{9}h_A(b_3+b_8)\,,\ \bar c_4=c_4-\frac{2}{9}h_A(b_3+b_8)\, .
\label{eqn:redefh}
\end{split}
\end{equation}
Here, the barred letters stand for the redefined constants.

\begin{figure}[tb]
\includegraphics[width=0.8\textwidth]{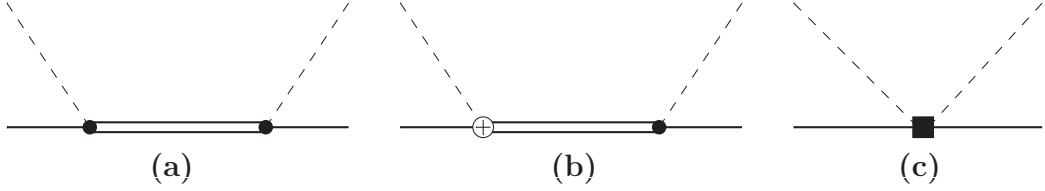}
\caption{Feynman diagrams showing the effect of $b_3+b_8$ $\pi N\Delta$ coupling on $\pi N$ scattering.
Dots denote the leading $\pi N\Delta$ vertex, the crossed circle stands for the $b_3+b_8$ vertex, and the filled square
denotes the subleading $\pi \pi NN$ contact term. Crossed diagrams are not shown.
}\label{fig1}
\end{figure}

As shown in \reffig{fig1}, the diagrammatic interpretation of eliminating $b_3 + b_8$ is that the subleading (with one vertex being $b_3+b_8$) $\Delta$ pole term (b) can be dissected to the sum of, up to some constant factors,
the leading $\Delta$ pole term (a) and the subleading $\pi\pi NN$ contact terms (c).
This can also be manifested by the identities
\beq
\frac{\omega}{\omega\pm\delta}=1\mp\frac{\delta}{\omega\pm\delta}\,,
\label{eqn:deltaprop}
\eeq
with the lower signs corresponding to pole diagrams while the upper ones to crossed.

Ref.~\cite{Krebs:2007rh}, in which $b_3 + b_8$ was employed, obtained two different sets of LEC values in fitting to $\pi N$ scattering data (note that our $h_A$ corresponds to $2h_A$ in their notation).
We calculate the LECs redefined according to Eq.~\eqref{eqn:redefh} and find that the two sets of LECs in Ref.~\cite{Krebs:2007rh} indeed correspond to the same set of barred LECs,
shown in \reftable{table:cis}, with only $\bar h_A$ having $\sim10\%$ discrepancy that can be traced to higher-order contributions.

\begin{table}[htb]
\begin{tabular*}{0.8\textwidth}{@{\extracolsep{\fill}}||l c|c|c|c|c|c||c|c|c|c||}
\hline
 &Source                         & $h_A$              & $b_3 + b_8$         &$c_2$               & $c_3$   & $c_4$              & $\bar h_A$         & $\bar c_2$ & $\bar c_3$         & $\bar c_4$ \\
\hline
 &Ref.~\cite{Krebs:2007rh} fit 1 & $\hphantom{-}2.68$ & $\hphantom{-}1.40$  & $-0.25$            & $-0.79$ & $\hphantom{-}1.33$ & $\hphantom{-}3.50$ & $-1.92$    & $\hphantom{-}0.88$ & $\hphantom{-}0.50$ \\
\hline
 &Ref.~\cite{Krebs:2007rh} fit 2 & $\hphantom{-}2.10$ & $\hphantom{-}2.95$  & $\hphantom{-}0.83$ & $-1.87$ & $\hphantom{-}1.87$ & $\hphantom{-}3.83$ & $-1.92$    & $\hphantom{-}0.88$ & $\hphantom{-}0.49$ \\
\hline
\end{tabular*}
\caption{Values of redefined LECs $\bar h_A$
(dimensionless) and $\bar c_2$, $\bar c_3$, $\bar c_4$ (in units of GeV$^{-1}$).
We also give the values of the input LECs (in the same respective units),
taken from fit~1 and fit~2 of Ref.~\cite{Krebs:2007rh}. Note that our $h_A$ corresponds to $2h_A$ in the notation of that reference. }\label{table:cis}
\end{table}

The presence of $b_3+b_8$ term in the Lagrangian at order $\mathcal{O}(p^2)$ can be a source of rather intricate troubles in certain calculations. Consider, for instance, a version of counting that does not go
together with the chiral index.
In this case the contributions due to $\mathcal{L}_{\pi N \Delta}^{b_3+b_8}$ can be of different (higher) order
than those coming from the $\pi\pi NN$ contact term with $c_i$'s. Examples of such countings are the $\delta$~counting,
used for calculations in the energy region extending to delta resonance~\cite{pascalutsa},
or the $p$~counting used in in calculations of the reactions $NN\to NN\pi$~\cite{NNpi}, where $\mathcal{L}_{\pi N \Delta}^{b_3+b_8}$
starts to contribute one order higher than the $\pi\pi NN$ contact term with $c_i$'s. This is due to the
nucleon-delta mass difference $\delta$ being considered as an intermediate scale ($\omega\ll\delta\ll \mhep$)
in these countings, and the ratio $\omega/\delta$ being just one of the expansion parameters,
which implies that one should expand the product of pion energy and the delta propagator in powers of $\omega/\delta$,
\beq
\frac{\omega}{\omega\pm\delta}=\pm\frac{\omega}{\delta}\left[1\mp\frac{\omega}{\delta}+\cdots\right]\,,
\label{eqn:deltapropNN}
\eeq
rather than use \refeq{eqn:deltaprop}. Hence, the contribution of $\mathcal{L}_{\pi N \Delta}^{b_3+b_8}$ starts one order higher than that of the terms with $c_i$'s.

A problem emerges when one attempts to use the values of $c_i$'s, calculated in an SSE calculation of $\pi N$ scattering,
in a calculation of $NN\to NN\pi$ up to next-to-next-to-leading order (NNLO), the order where $c_i$'s start to contribute.
The discussed correlations of $h_A$, $b_3+b_8$, and $c_i$'s [\refeq{eqn:redefh}] can lead to sizable variations of the latter, as illustrated by \reftable{table:cis},
thus leading to unnaturally large variations of the calculated observables in $NN\to NN\pi$. To counter these variations at NNLO, one would need to include
$\mathcal{L}_{\pi N \Delta}^{b_3+b_8}$, being one order higher according to $p$~counting. Such a promotion without a good reason would be completely undesired,
especially given the fact that there is a far more natural solution, namely, demoting the redundant $b_3+b_8$ term at the level of the Lagrangian, using the equations of motion,
and refitting $c_i$'s accordingly.

\section{Calculation of FW transformation}\label{app:details}
The FW transformation for a spin-$s$ field [Eq.~\eqref{eqn:Transform}] can be written in a closed analytical form as
\beq
F^{(s)}(\vec\nabla)=C\omega^{-1/2}\exp\left[\mp\left(\vec{\Sigma}^{(s)}\cdot\vec\nabla\right)\frac{1}{|\vec\nabla|}\arctanh-{i}\frac{|\vec\nabla|}{\omega}\right]\,,
\eeq
where $C$ is a normalization constant (see Ref.~\cite{PhysRev.135.B241}; note that we use a normalization of FW transformation that differs from one used in that reference).
In practice, it is convenient to express this exponential as a finite sum of terms constructed 
of the generators $\vec{\Sigma}^{(s)}$ and their products, as in Eq.~\eqref{eqn:Transform}. It is always possible due to the fact that
a product of any $2s+1$ generators $\vec{\Sigma}^{(s)}$ is a linear combination of products of up to $2s$ generators, which is manifested by the identity
\beq
\left[\left(\vec\Sigma^{(s)}\cdot\vec n\right)+s\right]\left[\left(\vec\Sigma^{(s)}\cdot\vec n\right)+(s-1)\right]\dots\left[\left(\vec\Sigma^{(s)}\cdot\vec n\right)-s\right]=0\, ,
\eeq
with $\vec n$ being an arbitrary unit vector. On the other hand, Eq.~\eqref{eqn:Transform} is the most general form of a transformation
of a spin-$s$ field conserving the rotational properties of that field, and one can use this equation as the starting point in deriving the FW transformation.

Demanding that the field $\Phi^{(s)}(x)$ [Eq.~\eqref{eqn:Transform}] transforms under the boost as given by Eq.~\eqref{eqn:SSpinors} and taking into account
the transformation properties of the gradient [Eq.~\eqref{eqn:transd}], the boost of the field $\chi_s(x)$ [Eq.~\eqref{eqn:transchi}], we arrive at the following equation for function
$F^{(s)}(\vec y)$:
\bea
F^{(s)}(\vec y)\left[\frac{{i}\vec y}{2\omega(y)}+\frac{\vec\Sigma^{(s)}\times \vec y}{m+\omega(y)}\right]-{i}\frac{\partial F^{(s)}(\vec y)}{\partial \vec y}\,\omega(y)=\pm\vec\Sigma^{(s)}F^{(s)}(\vec y)\,,
\label{eqn:MasterEquation}
\eea
where $\omega(y) = \sqrt{m^2 - y^2}$ and the choice of the sigh in the r.h.s.\ is the same as in Eq.~\eqref{eqn:SSpinors}. Here we again dropped terms proportional to the 
free EOM for the field $\chi_s(x)$. 
With the help of algebraic relations between the spin-$s$ generators $\Sigma^{(s)}_i$ and their products, one can
simplify this equation, arriving at a system of differential equations for functions  $f_i({\vec y}^{\,2})$.
In the following, we will show how this procedure works on the examples of spin-$0$, -$1/2$, and -$3/2$ fields.
Note that the transformation for the left-handed spinor, $F_L^{(s)}(\grad)$, corresponding to the positive sign in the
r.h.s.\ above, is related (up to normalization factors) to that for the right-handed spinor via the change of the sign of the gradient,
$F_R^{(s)}(\grad)=F_L^{(s)}(-\grad)$; hence, we will consider only the left-handed transformation below in this appendix.

\subsection{Spin-$0$}

For the spin-$0$ FW field $\chi_0(x)$ to remain a three-scalar, the field redefinition we are after must be
\beq
\Phi^{(0)}(x) = f_0(\vec{\nabla}^2) \chi_0(x) \, ,
\eeq
such that $\Phi^{(0)}(x)$ is a four-scalar, namely, $\Phi^{(0)} \to \Phi^{(0)}$ under the boost. Equation \eqref{eqn:MasterEquation} takes the form
\beq
f_0(y^2)\frac{\vec y\,}{2 \omega(y)} - \omega(y) \frac{\partial f_0(y^2)}{\partial \vec{y}} = 0 \, ,
\eeq
which gives, up to a normalization factor, $f_0(y^2)=\omega^{-1/2}$. One can choose, e.g.,
\beq
\Phi^{(0)}(x) = \sqrt{\frac{m}{\omega}}\,\chi_0(x) \, . \label{eqn:spin0}
\eeq

\subsection{Spin-$1/2$}

In the spin-$1/2$ case, we have the transformation for the (left-handed) spinors
\beq
\Phi^{(1/2)}_L = \left[f_0(\vec{\nabla}^2) + f_1(\vec{\nabla}^2) \vec{\sigma}\cdot\vec\nabla \right] \chi_\frac{1}{2} \, .
\eeq
Similar to the spin-$0$ case, one finds, substituting the expression for $\Phi^{(1/2)}$ in \eqref{eqn:MasterEquation} and using the algebra of the Pauli matrices,
\beq
\begin{split}
&\sigma_j\left[\frac{\epsilon_{ijk} y_k}{2(m+\omega)}f_0(\omega)+{i}\left(\frac{y_i y_j}{2\omega}+\frac{y_i y_j}{2(m+\omega)}-\frac{y^2\delta_{ij}}{2(m+\omega)}-\omega\delta_{ij}\right)f_1(\omega)
+{i} y_i y_j f_1^\prime(\omega)\right]\\
&+{i}\left(\frac{1}{2\omega}f_0(\omega)+f_0^\prime(\omega)\right)y_i\\
&=\frac{1}{2}\sigma_j\left[\delta_{ij}f_0(\omega)-{i}\epsilon_{ijk}y_k\,f_1(\omega)\right]+\frac{1}{2}y_i\, f_1(\omega)\,,
\end{split}
\eeq
where the functions $f_0$ and $f_1$ are considered as functions of $\omega$, and the prime denotes the derivative with respect to $\omega$. Considering factors in front of different
matrix and tensor structures appearing in this equation gives four equations: two differential equations and two algebraic equations relating $f_0$ and $f_1$.
One can choose any two of the four equations to solve for $f_0$ and $f_1$, for instance, the factors in front of $\epsilon_{ijk}\sigma_j y_k$ and $\sigma_jy_iy_j$ give
\beq
\begin{split}
\frac{f_0(\omega)}{m+\omega}&=-{i}f_1(\omega)\,,\\
\left(\frac{1}{2\omega}+\frac{1}{2(m+\omega)}\right)f_1(\omega)+f_1^\prime(\omega)&=0\,,
\end{split}
\eeq
which finally gives (up to normalization)
\beq
f_0(\omega)=\sqrt{\frac{m+\omega}{4\omega}}\,,\qquad f_1(\omega)=\frac{{i}}{\sqrt{4\omega(m+\omega)}}\,.
\eeq
The remaining two of the four equations are consistent with this solution for $f_0$ and $f_1$, and we arrive at the FW transformation for
the left-handed spin-$1/2$ spinor,
\beq
\Phi^{(1/2)}_L(x)=\sqrt{\frac{m+\omega}{4\omega}}\left[1+{i}\frac{(\vec\sigma\cdot\vec\nabla)}{m+\omega}\right]\chi_\frac{1}{2}(x)\,.
\eeq

\subsection{Spin-$3/2$}

The calculations for spin-$3/2$ are more involved due to the more complex algebra of the spin-$3/2$ generators, however, the ideology is the same as for spin-$1/2$.
It is convenient to introduce symmetrized versions of products of $\Sigma_i$ (here we again suppress the index $^{(3/2)}$ for the spin-$3/2$ generators):
\beq
\Sigma_{ijkl}=\frac{1}{24}\left\{\Sigma_i\Sigma_j\Sigma_k\Sigma_l\right\}\,,\qquad\Sigma_{ijk}=\frac{1}{6}\left\{\Sigma_i\Sigma_j\Sigma_k\right\}\,,\qquad\Sigma_{ij}=\frac{1}{2}\left\{\Sigma_i\Sigma_j\right\}\,,
\eeq
where the braces stand for summations over all the permutations of Cartesian indices. Using the standard commutation relations, $\left[\Sigma_i,\Sigma_j\right]={i}\epsilon_{ijk}\Sigma_k$,
one can get the following algebraic identities for the products of generators:
\bea
\Sigma_i\Sigma_j &=& \Sigma_{ij}+\frac{{i}}{2}\epsilon_{ijk}\Sigma_k\,,\\
\Sigma_i\Sigma_j\Sigma_k &=& \Sigma_{ijk}+\frac{{i}}{2}\left(\Sigma_{jl}\epsilon_{ikl}+\Sigma_{il}\epsilon_{jkl}+\Sigma_{kl}\epsilon_{ijl}\right)
+\frac{1}{6}\left(\delta_{il}\delta_{jk}-2\delta_{jl}\delta_{ik}+\delta_{ij}\delta_{kl}\right)\Sigma_l\,,\\
\Sigma_i\Sigma_j\Sigma_k\Sigma_l&=&\Sigma_{ijkl}+\frac{{i}}{2}\left(
\Sigma_{ijm}\epsilon_{mkl}+\Sigma_{ikm}\epsilon_{mjl}+\Sigma_{ilm}\epsilon_{mjk}+\Sigma_{jkm}\epsilon_{mil}+\Sigma_{jlm}\epsilon_{mik}+\Sigma_{klm}\epsilon_{mij}
\right)\nonumber\\
&+&\frac{1}{6}\left(2\Sigma_{ij}\delta_{kl}-
\Sigma_{ik}\delta_{jl}+
2\Sigma_{il}\delta_{jk}-
4\Sigma_{jk}\delta_{il}-
\Sigma_{jl}\delta_{ik}+
2\Sigma_{kl}\delta_{ij}\right)\nonumber\\
&-&\frac{1}{4}\Sigma_{mn}\left(
\epsilon_{ijm}\epsilon_{kln}+
\epsilon_{ikm}\epsilon_{jln}+
\epsilon_{ilm}\epsilon_{jkn}
\right)
+\frac{{i}}{12}\left(
2\epsilon_{ijl}\Sigma_k+
2\epsilon_{jkl}\Sigma_i-
 \epsilon_{ikl}\Sigma_j-
3\epsilon_{ijk}\Sigma_l
\right)\nonumber\\
&-&\frac{{i}}{12}\left(
4\epsilon_{ikm}\delta_{jl}-
 \epsilon_{ijm}\delta_{kl}-
3\epsilon_{ilm}\delta_{jk}-
 \epsilon_{jkm}\delta_{il}+
2\epsilon_{jlm}\delta_{ik}+
 \epsilon_{klm}\delta_{ij}
\right)\Sigma_m\,.
\eea
The linear dependence of $\Sigma_{ijkl}$ on products of lower powers of $\Sigma_i$ is given by the identity
\beq
\Sigma_{ijkl}=\frac{5}{12}\left(
\Sigma_{ij}\delta_{kl}+
\Sigma_{ik}\delta_{jl}+
\Sigma_{il}\delta_{jk}+
\Sigma_{jk}\delta_{il}+
\Sigma_{jl}\delta_{ik}+
\Sigma_{kl}\delta_{ij}
\right)-\frac{3}{16}\left(
\delta_{ij}\delta_{kl}+
\delta_{ik}\delta_{jl}+
\delta_{il}\delta_{jk}
\right)\,.
\eeq

We write the transformation for the left-handed four-component spinor as
\beq\label{eqn:3halftrans}
\Phi^{(3/2)}_L = \left[f_0(\vec{\nabla}^2) + f_1(\vec{\nabla}^2) \vec{\Sigma}\cdot\vec\nabla +f_2(\vec{\nabla}^2)M_{ij}\nabla_i\nabla_j+ f_3(\vec{\nabla}^2)T_{ijk}\nabla_i\nabla_j\nabla_k\right] \chi_\frac{3}{2} \, ,
\eeq
where matrices $M_{ij}$ and $T_{ijk}$ are defined in Eq.~\eqref{eqn:MTdef}:
\beq\nonumber
M_{ij}=\Sigma_{ij}-\frac{5}{4}\delta_{ij},\quad T_{ijk}=\Sigma_{ijk}-\frac{41}{60}\left(\Sigma_i\delta_{jk}+\Sigma_j\delta_{ik}+\Sigma_k\delta_{ij}\right)\,.
\eeq
This definition is to ensure that a contraction of the two Cartesian indices of $M_{ij}$ (or any two indices of $T_{ijk}$) gives zero result,
resulting in the set of quantities $\chi^\dagger_\frac{3}{2} M_{ij} \chi_\frac{3}{2}$ being an SO(3) irreducible tensor of rank two,
i.e.\ it transforms under rotations under a spin-$2$ irreducible representation of SO(3), because lower spin (spin-$0$ in this specific case) components
are proportional to the Cartesian trace of this tensor and therefore vanish by the definition of $M_{ij}$. Analogously, $\chi^\dagger_\frac{3}{2} T_{ijk} \chi_\frac{3}{2}$
is an SO(3) irreducible tensor of rank three, having only spin-$3$ components.

Plugging the transformation for $\Phi^{(3/2)}$ into Eq.~\eqref{eqn:MasterEquation} and using the spin-$3/2$ algebra given above, one can
obtain equations for the four functions $f_0(\omega),\,\dots\,,f_3(\omega)$, in full analogy to the spin-$0$ and spin-$1/2$ cases considered before.
The calculation is, however, much more tedious due to the convoluted algebra, and was done in practice with help of symbolic calculation software~\cite{FORM},
with the final result for $\Phi^{(3/2)}_L$ given by Eq.~\eqref{eqn:zetalr} in the main text, and repeated here for the sake of completeness:
\beq\nonumber
\Phi^{(3/2)}_L(x) = \sqrt{\frac{m+\omega}{4\omega}} \left[\frac{\omega}{m} +{i}\frac{(6\omega+4m)\vec\Sigma\cdot\vec\nabla}{5m(m+\omega)}-\frac{M_{ij}\nabla_i\nabla_j}{m(m+\omega)}
-{i}\frac{2T_{ijk}\nabla_i\nabla_j\nabla_k}{3m(m+\omega)^2}\right] \chi_\frac{3}{2}(x) \, .
\eeq

\section{$1/m_N$ expansion of $NN$ and $N\Delta$ bilinears}\label{app:exp}
In this Appendix we give $1/m_N$ expansion of covariant nucleon-nucleon and nucleon-delta bilinears, in terms of heavy fields $N$ and $\Delta$, as defined in Eq.~\eqref{eqn:defND}.
This is a straightforward computation of Tables~\ref{tbl:bihalf} and \ref{tbl:bi3half}, with Eqs.~\eqref{eqn:etalr}, \eqref{eqn:zetalr}, and \eqref{eqn:defND} applied.

\subsection{$NN$ bilinears}

\begin{itemize}
 \item Scalar $s = \overline{\psi}_D \psi_D$ :
 \bea
 s = 
N^\dagger
\left[1 + \frac{(\loarrow\nabla-\roarrow\nabla)^2}{8m_N^2}
-i\frac{\loarrow\nabla\times\roarrow\nabla\cdot\vec\sigma}{4m_N^2} \right]
N
 \, .
 \eea
 \item Pseudo-scalar $p = \overline{\psi}_D \gamma^5 \psi_D$ :
 \bea
 p = 
N^\dagger
\left[-i\frac{(\loarrow\nabla+\roarrow\nabla)\cdot\vec\sigma}{2m_N}\right]
N \, . \label{eqn:NNp}
 \eea
 \item Vector $v^\mu = \overline{\psi}_D \gamma^\mu \psi_D$ :
 \bea
 v^0
 &=&
N^\dagger
\left[1 + \frac{(\loarrow\nabla+\roarrow\nabla)^2}{8m_N^2}
+ i\frac{\loarrow\nabla\times\roarrow\nabla\cdot\vec\sigma}{4m_N^2}\right]
N \, ,\nonumber \\
 \vec v
&=&
N^\dagger
\left[i\frac{(\loarrow\nabla-\roarrow\nabla)}{2m_N} + \frac{(\loarrow\nabla+\roarrow\nabla)\times\vec\sigma}{2m_N}\right]
N \, \label{eqn:NNv}.
 \eea
 \item Pseudo-vector $a^\mu = \overline{\psi}_D \gamma^5 \gamma^\mu \psi_D$ :
 \bea
 a^0
 &=&
N^\dagger\left[ -i\frac{(\loarrow\nabla-\roarrow\nabla)\cdot\vec\sigma}{2m_N}\right]
N \, , \nonumber \\
 \vec a
 &=&
N^\dagger
\left[-\vec\sigma -\frac{(\loarrow\nabla\cdot\vec\sigma)\roarrow\nabla+\loarrow\nabla(\roarrow\nabla\cdot\vec\sigma)}{4m_N^2}
- \frac{\vec\sigma( \loarrow\nabla - \roarrow\nabla )^2}{8m_N^2} 
+i\frac{\loarrow\nabla\times\roarrow\nabla}{4m_N^2}\right]
N \, . \label{eqn:NNa}
 \eea
 \item Tensor $F^{\mu \nu} = \overline{\psi}_D \sigma^{\mu \nu} \psi_D$ :
\bea
 F^{0i}
&=&
N^\dagger
\left[\frac{\loarrow\nabla+\roarrow\nabla}{2 m_N}
- i\frac{(\loarrow\nabla-\roarrow\nabla)\times\vec\sigma}{2m_N}\right]_i
N \, , \nonumber \\
F^{ij}
&=&
N^\dagger
\left[\epsilon_{ijk}\sigma_k
+i\frac{\loarrow\nabla_i\roarrow\nabla_j-\loarrow\nabla_j\roarrow\nabla_i}{4m_N^2}
+\frac{\epsilon_{ikl}\loarrow\nabla_k\sigma_l\roarrow\nabla_j-\epsilon_{jkl}\loarrow\nabla_k\sigma_l\roarrow\nabla_i}{4m_N^2}\right.\nonumber\\
& & {}-\left.\frac{(\loarrow\nabla\cdot\vec\sigma)\epsilon_{ijk}\roarrow\nabla_k}{4m_N^2}
+\frac{\epsilon_{ijk}\sigma_k(\loarrow\nabla^2+\roarrow\nabla^2)}{8m_N^2} \right]
N \,
 \label{eqn:NNF}.
 \eea
\end{itemize}

\subsection{$N\Delta$ bilinears}

\begin{itemize}
 \item $G^{\mu\nu}$:
 \bea
 G^{0i}
&=&
N^\dagger
\left[S_i+\frac{(\loarrow\nabla\cdot\vec S)\roarrow\nabla_i+\loarrow\nabla_i(\vec S\cdot\roarrow\nabla)}{8m_N^2}+\frac{3(\vec S\cdot\roarrow\nabla)\roarrow\nabla_i}{4m_N^2}
-\frac{S_i}{8m_N^2}(5\roarrow\nabla^2-\loarrow\nabla^2+4\loarrow\nabla\cdot\roarrow\nabla)\right.\nonumber\\
&-&\left.i\frac{\loarrow\nabla_l\roarrow\nabla_j+\roarrow\nabla_l\roarrow\nabla_j}{2m_N^2}\Omega_{kl}\epsilon_{ijk}
+i\frac{\loarrow\nabla_l\roarrow\nabla_j}{4m_N^2}\Omega_{ik}\epsilon_{jlk}\right]
\Delta\, ,\nonumber\\
 G^{ij}
&=&
N^\dagger
\left[i\frac{S_i}{4m_N} ( \loarrow\nabla_j + 5\roarrow\nabla_j- \frac{5\delta}{m_N}\roarrow\nabla_j)
       - i\frac{S_j}{4m_N}( \loarrow\nabla_i+5\roarrow\nabla_i -\frac{5\delta}{m_N}\roarrow\nabla_i)
\vphantom{\frac{\delta\roarrow\nabla_k}{2m_N^2}}
\right.\nonumber\\
&-&\left.\frac{\loarrow\nabla_k+\roarrow\nabla_k}{2m_N}\Omega_{kl}\epsilon_{ijl}+\frac{\delta\roarrow\nabla_k}{2m_N^2}\Omega_{kl}\epsilon_{ijl}\right]
\Delta\, .
\label{eqn:NDG}
 \eea
\item $F^{\mu\nu\lambda}$ :
\bea
F^{00i}
 &=&
N^\dagger
\left[S_i-\frac{(\loarrow\nabla\cdot\vec S)\roarrow\nabla_i+\loarrow\nabla_i(\vec S\cdot\roarrow\nabla)}{8m_N^2} +\frac{3(\vec S\cdot\roarrow\nabla)\roarrow\nabla_i}{4m_N^2}
-\frac{S_i}{8m_N^2}(5\roarrow\nabla^2-\loarrow\nabla^2-4\loarrow\nabla\cdot\roarrow\nabla)\right.\nonumber\\
&+&\left.i\frac{\loarrow\nabla_j\roarrow\nabla_l -\roarrow\nabla_j\roarrow\nabla_l}{2m_N^2}\Omega_{jk}\epsilon_{ilk}
 - i\frac{\loarrow\nabla_l\roarrow\nabla_j}{4m_N^2}\Omega_{ik}\epsilon_{jlk}\right]
\Delta \, ,\nonumber \\
F^{0ij}
&=&
N^\dagger
\left[
- i \frac{S_i}{4m_N}  ( \loarrow\nabla_j - 5\roarrow\nabla_j +\frac{5\delta}{m_N}\roarrow\nabla_j)
+ i \frac{S_j}{4m_N}  ( \loarrow\nabla_i - 5\roarrow\nabla_i +\frac{5\delta}{m_N}\roarrow\nabla_i)
\vphantom{\frac{\delta\roarrow\nabla_k}{2m_N^2}}
\right.\nonumber\\
&+&
\left.\frac{\loarrow\nabla_k-\roarrow\nabla_k}{2m_N}\Omega_{kl}\epsilon_{ijl}+\frac{\delta\roarrow\nabla_k}{2m_N^2}\Omega_{kl}\epsilon_{ijl}\right]
\Delta \, ,\nonumber \\
F^{ij0}
&=&
N^\dagger
\left[-i\frac{S_i}{4m_N}(\loarrow\nabla_j+\roarrow\nabla_j-\frac{\delta}{m_N}\roarrow\nabla_j)
 -i\frac{S_j}{2m_N}(\loarrow\nabla_i-2\roarrow\nabla_i+\frac{2\delta}{m_N}\roarrow\nabla_i)+i\frac{(\loarrow\nabla\cdot\vec S)-(\roarrow\nabla\cdot\vec S)}{4m_N}\delta_{ij}\right.\nonumber\\
&+&\left.i\frac{\delta(\loarrow\nabla\cdot\vec S)}{4m_N^2}\delta_{ij}-\frac{\roarrow\nabla_k}{m_N}\Omega_{il}\epsilon_{jkl}-\frac{\loarrow\nabla_k+\roarrow\nabla_k}{2m_N}\Omega_{jl}\epsilon_{ikl}
+\frac{\delta\roarrow\nabla_k}{m_N^2}\Omega_{il}\epsilon_{jkl}+\frac{\delta\roarrow\nabla_k}{2m_N^2}\Omega_{jl}\epsilon_{ikl}\right]
\Delta \,,\nonumber \\
F^{ijk}
&=&
N^\dagger
\left[-i\Omega_{il}\epsilon_{jkl}-\frac{1}{2}S_j\delta_{ik}+\frac{1}{2}S_k\delta_{ij}
-\frac{(\loarrow\nabla\cdot\vec S)}{4m_N^2}(\roarrow\nabla_j\delta_{ik}-\roarrow\nabla_k\delta_{ij})
-\frac{S_i}{4m_N^2}(\loarrow\nabla_j\roarrow\nabla_k-\loarrow\nabla_k\roarrow\nabla_j)\right.\nonumber\\
&+&\frac{S_j}{16m_N^2}(-10\loarrow\nabla_i\roarrow\nabla_k-2\loarrow\nabla_k\roarrow\nabla_i+12\roarrow\nabla_i\roarrow\nabla_k-(\loarrow\nabla-\roarrow\nabla)^2\delta_{ik}
)
\nonumber\\
&-&\frac{S_k}{16m_N^2}(-10\loarrow\nabla_i\roarrow\nabla_j-2\loarrow\nabla_j\roarrow\nabla_i+12\roarrow\nabla_i\roarrow\nabla_j-(\loarrow\nabla-\roarrow\nabla)^2\delta_{ij}
)
\nonumber\\
&-&\frac{\loarrow\nabla_m\roarrow\nabla_n}{4m_N^2}S_l\epsilon_{jkl}\epsilon_{imn}
-i\frac{\loarrow\nabla_m\roarrow\nabla_i+\roarrow\nabla_m\loarrow\nabla_i}{4m_N^2}\Omega_{ml}\epsilon_{jkl}
+i\frac{\roarrow\nabla_m\roarrow\nabla_k}{2m_N^2}\Omega_{il}\epsilon_{jml}-i\frac{\roarrow\nabla_m\roarrow\nabla_j}{2m_N^2}\Omega_{il}\epsilon_{kml}\nonumber\\
&-&\left. i\frac{(\loarrow\nabla-\roarrow\nabla)^2}{8m_N^2}\Omega_{il}\epsilon_{jkl}
+
i\frac{\loarrow\nabla_m\roarrow\nabla_k+\roarrow\nabla_m\roarrow\nabla_k}{2m_N^2}\Omega_{jl}\epsilon_{iml}
-i\frac{\loarrow\nabla_m\roarrow\nabla_j+\roarrow\nabla_m\roarrow\nabla_j}{2m_N^2}\Omega_{kl}\epsilon_{iml}\right]
\Delta \,.
\label{eqn:NDF}
 \eea

\item $H^{\mu\nu\lambda\rho}$ :
\beq
\begin{split}
H^{0i0j}
 &=
N^\dagger
\left[
i\frac{(\loarrow\nabla\cdot\vec S)+(\roarrow\nabla\cdot\vec S)}{4m_N}\delta_{ij}-i\frac{\delta(\roarrow\nabla\cdot\vec S)}{4m_N^2}\delta_{ij}\right.\\
&-\left.i\frac{3S_i}{8m_N}(\loarrow\nabla_j+\roarrow\nabla_j-\frac{\delta}{m_N}\roarrow\nabla_j)-i\frac{3S_j}{8m_N}(\loarrow\nabla_i+\roarrow\nabla_i-\frac{\delta}{m_N}\roarrow\nabla_i)\right.\\
&-\left.\frac{\Omega_{in}}{4m_N}\epsilon_{jrn}(\loarrow\nabla_r-3\roarrow\nabla_r+\frac{3\delta}{m_N}\roarrow\nabla_r)
-\frac{\Omega_{jn}}{4m_N}\epsilon_{irn}(\loarrow\nabla_r-3\roarrow\nabla_r+\frac{3\delta}{m_N}\roarrow\nabla_r)
\vphantom{\frac{(\loarrow\nabla\cdot\vec S)+(\roarrow\nabla\cdot\vec S)}{4m_N}}
 \right]
\Delta \, ,\\
H^{0ijk}
&=
N^\dagger
\left[-i\Omega_{in}\epsilon_{jkn}
+\frac{1}{8m_N^2}((\loarrow\nabla\cdot\vec S)(\roarrow\nabla_j\delta_{ik}-\roarrow\nabla_k\delta_{ij})
               -(\roarrow\nabla\cdot\vec S)(\loarrow\nabla_j\delta_{ik}-\loarrow\nabla_k\delta_{ij}))\right.\\
&+\left.
\frac{S_j}{16m_N^2}(5\loarrow\nabla_i\roarrow\nabla_k+\loarrow\nabla_k\roarrow\nabla_i+6\roarrow\nabla_i\roarrow\nabla_k)
-\frac{S_k}{16m_N^2}(5\loarrow\nabla_i\roarrow\nabla_j+\loarrow\nabla_j\roarrow\nabla_i+6\roarrow\nabla_i\roarrow\nabla_j)\right.\\
&+\left.\frac{S_n}{16m_N^2}\left(\loarrow\nabla_r\roarrow\nabla_s(\epsilon_{irs}\epsilon_{jkn}-\epsilon_{irn}\epsilon_{jks}-5\epsilon_{isn}\epsilon_{jkr})-6\roarrow\nabla_r\roarrow\nabla_s\epsilon_{irn}\epsilon_{jks}\right)
\right.\\
&+\left.\frac{3S_i}{8m_N^2}(\loarrow\nabla_j\roarrow\nabla_k-\loarrow\nabla_k\roarrow\nabla_j)
+i\frac{\Omega_{in}}{8m_N^2}(\loarrow\nabla_n\roarrow\nabla_r\epsilon_{jkr}+\loarrow\nabla_r\roarrow\nabla_n\epsilon_{jkr}-2\roarrow\nabla_n\roarrow\nabla_r\epsilon_{jkr})
\right.\\
&+\left.i\frac{\Omega_{rn}}{8m_N^2}\epsilon_{jkn}(\loarrow\nabla_i\roarrow\nabla_r+\loarrow\nabla_r\roarrow\nabla_i-2\roarrow\nabla_r\roarrow\nabla_i)
+i\frac{\Omega_{in}}{8m_N^2}\epsilon_{jkn}(3\roarrow\nabla^2-\loarrow\nabla^2-2\loarrow\nabla\cdot\roarrow\nabla)\right.\\
&+\left.i\frac{\Omega_{jn}}{4m_N^2}\epsilon_{irn}(2\roarrow\nabla_r\roarrow\nabla_k-\loarrow\nabla_r\roarrow\nabla_k-\loarrow\nabla_k\roarrow\nabla_r)
-i\frac{\Omega_{kn}}{4m_N^2}\epsilon_{irn}(2\roarrow\nabla_r\roarrow\nabla_j-\loarrow\nabla_r\roarrow\nabla_j-\loarrow\nabla_j\roarrow\nabla_r)
 \right]
\Delta \, ,\\
H^{ijkl}
&=
N^\dagger
\left[
 i\frac{(\loarrow\nabla\cdot\vec S)}{8m_N}(\delta_{il}\delta_{jk}-\delta_{ik}\delta_{jl})
-i\frac{(\roarrow\nabla\cdot\vec S)}{8m_N}\left(1-\frac{\delta}{m_N}\right)(\delta_{il}\delta_{jk}-\delta_{ik}\delta_{jl})\right.\\
&+\left.i\frac{S_i}{4m_N}\left(1-\frac{\delta}{m_N}\right)(\roarrow\nabla_l\delta_{jk}-\roarrow\nabla_k\delta_{jl})
-i\frac{S_j}{4m_N}\left(1-\frac{\delta}{m_N}\right)(\roarrow\nabla_l\delta_{ik}-\roarrow\nabla_k\delta_{il})\right.\\
&+\left.i\frac{S_k}{8m_N}\left(\loarrow\nabla_j\delta_{il}-\loarrow\nabla_i\delta_{jl}+(\roarrow\nabla_j\delta_{il}-\roarrow\nabla_i\delta_{jl})\left(1-\frac{\delta}{m_N}\right)\right)\right.\\
&-\left.i\frac{S_l}{8m_N}\left(\loarrow\nabla_j\delta_{ik}-\loarrow\nabla_i\delta_{jk}+(\roarrow\nabla_j\delta_{ik}-\roarrow\nabla_i\delta_{jk})\left(1-\frac{\delta}{m_N}\right)\right)\right.\\
&+\left.i\frac{S_n}{8m_N}\left(3\loarrow\nabla_r\epsilon_{ijr}\epsilon_{kln}+\roarrow\nabla_r\epsilon_{ijr}\epsilon_{kln}\left(1-\frac{\delta}{m_N}\right)\right)
\right. \\
&+ \left. 
i\frac{S_n}{4m_N}\left(\loarrow\nabla_r\epsilon_{klr}\epsilon_{ijn}+\roarrow\nabla_r\epsilon_{klr}\epsilon_{ijn}\left(1-\frac{\delta}{m_N}\right)\right)\right.\\
&-\left.\frac{1}{4m_N}\Omega_{in}\epsilon_{kln}\left(\loarrow\nabla_j-3\roarrow\nabla_j\left(1-\frac{\delta}{m_N}\right)\right)
+\frac{1}{4m_N}\Omega_{jn}\epsilon_{kln}\left(\loarrow\nabla_i-3\roarrow\nabla_i\left(1-\frac{\delta}{m_N}\right)\right)\right.\\
&-\left.\frac{1}{4m_N}\Omega_{kn}\epsilon_{ijn}\left(\loarrow\nabla_l-3\roarrow\nabla_l\left(1-\frac{\delta}{m_N}\right)\right)
+\frac{1}{4m_N}\Omega_{ln}\epsilon_{ijn}\left(\loarrow\nabla_k-3\roarrow\nabla_k\left(1-\frac{\delta}{m_N}\right)\right)
\vphantom{\frac{(\loarrow\nabla\cdot\vec S)}{8m_N}}
 \right]
\Delta \,.
\end{split}
\label{eqn:NDH}
\eeq
\end{itemize}

\end{document}